\documentclass[5p,authoryear]{elsarticle}
  
\journal{Eur.~J.~Mech.~A}
\usepackage{hyperref}
\usepackage{xcolor}
  \hypersetup{colorlinks,%
    citecolor=blue,%
    filecolor=blue,%
    linkcolor=blue,%
    urlcolor=blue}

\usepackage[utf8]{inputenc}
\usepackage{graphicx}
\usepackage{amsmath,amssymb,mathtools}
\usepackage{tikz}
\usetikzlibrary{shapes,arrows,positioning}
\usepackage{enumitem,setspace}
\usepackage{array}
\usepackage{pifont}
\usepackage{lmodern} 

\newcommand{\intd}{\mathrm{d}} 
\newcommand{\pdr}[2]{\frac{\partial {#1}}{\partial {#2}}} 
\newcommand{\ppdr}[2]{\frac{\partial^2 {#1}}{\partial {#2}^2}} 
\newcommand{\myquad}[1][1]{\hspace*{#1em}\ignorespaces}
\newcommand{\vect}[1]{\boldsymbol{#1}}
\newcommand{\Abaqus}{\textsc{Abaqus}} 
\newcommand{\Matlab}{\textsc{Matlab}} 
\newcommand{\intA}{{A^\wedge}} 
\newcommand{\intB}{{B^\wedge}} 
\DeclareMathOperator{\sgn}{sgn}
\newcommand{\quotes}[1]{``#1''}
\newcommand{\cmark}{\ding{51}} 
\newcommand{\xmark}{\ding{55}} 

\def\clap#1{\hbox to 0pt{\hss#1\hss}}

\def\mathrlap{\mathpalette\mathrlapinternal}

\def\mathrlapinternal#1#2{%
\rlap{$\mathsurround=0pt#1{#2}$}}

\begin{document}

  \begin{frontmatter}
  \title{\textbf{Through-Thickness Modelling of Metal Rolling\\using Multiple-Scales Asymptotics}}

\author[wmi]{Mozhdeh~Erfanian}\ead{Mozhdeh.Erfanian@warwick.ac.uk}
\author[wmi,wmg]{Edward~J.~Brambley\corref{cor1}}\ead{E.J.Brambley@warwick.ac.uk}\cortext[cor1]{Corresponding Authors}
\author[macsi]{Francis~Flanagan}\ead{Francis.Flanagan@ul.ie}
\author[macsi]{Doireann~O'Kiely\corref{cor1}}\ead{Doireann.OKiely@ul.ie}
\author[csis,lero]{Alison~N.~O'Connor}\ead{Alison.OConnor@ul.ie}

 \affiliation[wmi]{organization={Mathematics Institute},
             addressline={University of Warwick},
             city={Coventry},
             postcode={CV4 7AL},
             country={UK}}

  \affiliation[wmg]{organization={WMG},
             addressline={University of Warwick},
             city={Coventry},
             postcode={CV4 7AL},
             country={UK}}

  \affiliation[macsi]{organization={MACSI, Department of Mathematics \& Statistics},
             addressline={University of Limerick},
             city={Limerick},
             postcode={V94 T9PX},
             country={Ireland}}

 \affiliation[csis]{organization={Computer Science and Information Systems (CSIS)},
             addressline={University of Limerick},
             city={Limerick},
             postcode={V94 T9PX},
             country={Ireland}}

  \affiliation[lero]{organization={Lero, The Science Foundation Ireland Centre for Software Research},
             addressline={University of Limerick},
             city={Limerick},
             postcode={V94~T9PX},
             country={Ireland}}

\def\elspublication{Published in European Journal of Mechanics --- A/Solids, volume~113, 105712 (2025), \href{https://doi.org/10.1016/j.euromechsol.2025.105712}{doi:10.1016/j.euromechsol.2025.105712}}

\begin{abstract}
A new semi-analytic model of the metal rolling process is introduced, which, for the first time, is able to predict the through-thickness stress and strain oscillations present in long thin roll-gaps.
The model is based on multiple-scales asymptotics, assuming a long thin roll-gap and a comparably small Coulomb friction coefficient.
The leading-order solution varies only on a long lengthscale corresponding to the roll-gap length and matches with slab models. 
The next-order correction varies on both this long lengthscale and a short lengthscale associated with the workpiece thickness, and reveals rapid stress and strain oscillation both in the rolling direction and through the thickness. 
For this initial derivation, the model assumes a rigid perfectly-plastic material behaviour.
Despite these strong assumptions, this model compares well with finite element simulations that employ more realistic material behaviour (including elasticity and strain hardening).
These assumptions facilitate the simplest possible model to provide a foundational understanding of the complex through-thickness behaviour observed in the finite element simulations, while requiring an order of only seconds to compute. 
This model can form the foundation of further improved models with more complicated mechanics in the future.
\Matlab\ code for evaluating the model is provided in the supplementary material.
\end{abstract}


\end{frontmatter}

\section{Introduction}

\begin{figure*}%
\centering%
\includegraphics[]{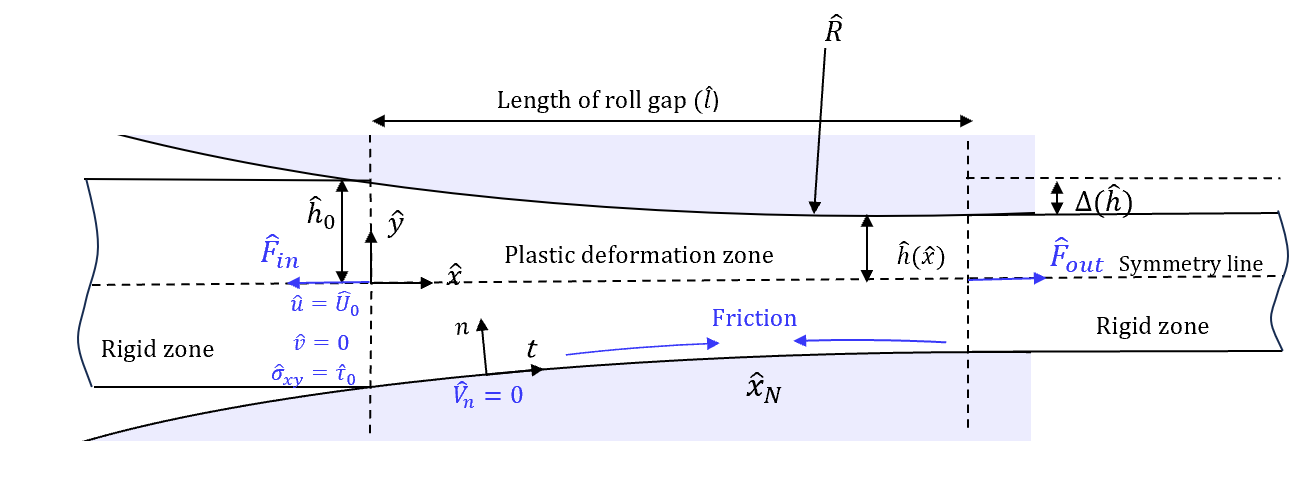}%
\caption{Schematic diagram of symmetric rolling with the domain of interest and the boundary conditions. The sheet, which is wide in the direction into the page, enters the roll gap between the two rolls from the left and exits as a thinner sheet to the right.}%
\label{rollgap}%
\end{figure*}

Rolling, depicted in Figure~\ref{rollgap}, is the process of reducing the thickness of a sheet (the workpiece) by passing it between two work rolls.  
It is a major industrial process in metal manufacturing, with over 99\% of cast steel and two-thirds of wrought aluminium being rolled~\cite[][pp.~54--55]{allwood+cullen-2012}. 
The rolling process is usually repeated for several
passes until the desired thickness is achieved, either by moving the workpiece backwards and forward through the same pair of rolls, or by having several roll stands in tandem which the workpiece passes through in sequence.  
One or both of the work rolls is mechanically driven, and the workpiece is pulled through the roll gap by friction between the workpiece and the rolls.  
In addition to altering the thickness, cold rolling alters the material's physical properties, which can be of more importance for the final product~\citep{lenard2013primer}.

The modelling of rolling is significant to industry as a potential means of reducing material waste, and driving towards carbon neutrality and sustainability~\citep{allwood2016closed}. 
Modern set-up and control algorithms for rolling mills aspire to go beyond controlling only geometry, utilizing models which, in addition to predicting how key quantities such as roll load and torque depend on the rolling parameters, also include basic microstructural modelling~\citep{idzik2024coupling}.
Finite Element (FE) simulations are considered the `gold standard' method of analysing rolling processes \citep{trzepiecinski2017effect,prabhu2020multi,kumar2021simulation}.
However, despite their accuracy, the long computational time of FE simulations prohibits their deployment in real-time applications.
Moreover, FE simulations of rolling have shown to not include sufficient through-thickness resolution to model fine through-thick\-ness variations until recently~\citep{flanagan2024}, which is a pre-requisite for accurately modelling microstructural development~\citep{decroos2014new}.
These limitations motivate the exploration of fast models that can support real-time process control while including through-thickness variation. 
Fast models are often based on equations drawn from physical principles, with some simplifying assumptions that enable analytical or semi-analytical solution. 

Historically, fast rolling models aimed to predict only roll force and roll torque in order to enable gauge control~\citep{von1925theory,siebel1927}. 
These fast models were established from the equilibrium of forces acting on each vertical slab of material in the zone of deformation: a method commonly known as the `slab method'. 
Slab models were subsequently improved by including more complex material models, and accounting for roll elasticity by adjusting the roll shape~\cite[e.g.][]{roychoudhuri1984mathematical,fleck1992cold}. 
However, the slab method is unable to predict through-thickness variation; in effect, the slab method involves an implicit assumption that the deformation is homogeneous through-thickness.  

\Citet{orowan1943calculation} was among the first to discard the assumption of homogeneous through-thickness deformation by modifying the shape of the slab to a thin segment bounded by circular arcs. 
The \citet{Nadai1931} solution is then applied to this new geometry, which enabled the inclusion of through-thickness shear in the model. 
A series of developments followed~\citep{Orowan1946first,bland1948calculation,bland1952approximate} which extended Orowan’s original work to incorporate tensions or to simplify the model with additional assumptions.

Models originating from a slab method predict a single pressure peak within the roll gap.
However, \citet{macgregor1959distribution} experimentally found a wave-like roll pressure with multiple peaks. 
Similarly, using the method of sliplines, for a sheet with comparable half thickness and length, \Citet{firbank1965suggested} also found two pressure peaks with a marked drop in pressure between them.
Subsequently, \citet{al1973experimental} conducted a series of experiments with different materials and aspect ratios and measured multiple pressure peaks. 
This oscillatory pattern was later verified using FE analyses~\citep{yarita1985stress,prakash1995steady,montmitonnet2006hot,flanagan2023,flanagan2024}, which demonstrates that the wave-like roll pressure observed is a consequence of a complicated through-thickness behaviour.
These findings imply that slab models, due to their simplified through-thickness assumptions, can not accurately model the wave-like oscillations noted to occur in both experiments and FE simulations.
It is clear that the oscillatory patterns are critical in understanding how the mechanical material behaviour and the cold rolling process parameters interact.
Without accurately capturing these patterns, the impact of any real-time monitoring algorithms will be severely limited.

A more rigorous mathematical framework for deriving fast approximate models is asymptotic analysis.
Asymptotics utilises systematic assumptions of scale, as opposed to ad-hoc assumptions of unknown error and limitation~\citep{minton2016asymptotic}. 
Asymptotics is especially effective in situations where parameters are not of the same magnitude (i.e., one parameter is notably smaller compared to the others). 
As such, asymptotics has found successful application in the modelling of rolling, where the presence of a small parameter is readily identifiable.
\Citet{smet1989asymptotic} developed the first asymptotic formulation of the sheet metal rolling process and accounted for through-thickness variation in the flow and the inhomogeneous work hardening it caused.
This technique has subsequently been used in other studies with different perturbation parameters, such as  the ratio of yield stress to maximum pressure by \citet{domanti1995two}, the ratio of entry-gauge thickness to roll-bite length initially by \citet{smelserjohnson1992asymptotic} and more recently for sandwich rolling~\citep{cawthorn2016asymptotic} and asymmetric clad-sheet rolling~\citep{minton2016asymptotic}. 
Although asymptotic models incorporate the effect of shear evolution through the thickness, they are often only validated at the workpiece surface, which does not necessarily validate the through-thickness heterogeneity \citep{smet1989asymptotic,domanti1995two,smelserjohnson1992asymptotic}.

FE simulations can be used to validate theoretical models.
In fact, validation via FE simulation is preferable to experimental methods given the difficulty and expense of experimentally measuring through-thickness stress and strain.  
However, comparisons of theoretical models with FE simulations are infrequent in the literature.  
\Citet{Mintonthesis} compared his asymptotic models for different rolling processes with results from the commercial FE package \Abaqus~\citep{Abaqus}. 
He reported through-thickness oscillatory patterns in the shear stress fields for all FE simulations of asymmetric, thick sheet and clad-sheet rolling. 
He demonstrated that the oscillatory pattern is pervasive in rolling, regardless of the material's rigidity or work hardening properties, the type of FE solver used (implicit or explicit), and whether or not plane strain is assumed.  
These oscillatory patterns, however, are not captured in his asymptotic models~\citep{cawthorn2016asymptotic,minton2016asymptotic}.  

Careful FE simulations~\citep{Mintonthesis,flanagan2023,flanagan2024} show that the through-thickness pattern oscillates rapidly in the axial direction on the length scale of the sheet thickness; a length scale which previous asymptotic analyses have ignored as small compared with the overall deformation.
The existence of effects on two different length scales in the same axial direction requires a different type of asymptotic analysis, termed multiple scales~\citep{hinch-ch7}, in which a fast scale provides rapid small variations about a solution which slowly evolves on the slow scale.  The slowly varying solution is sometimes referred to as a slow manifold, invariant manifold, or centre manifold, and is a general concept not unique to asymptotics~\citep[see, e.g.][]{roberts2015model}.  These techniques of a fast variation about a slowly-varying solution have been applied to elastic beams~\citep{valeryroy2002lubrication}, lubricating flows~\citep{roberts1993invariant}, and acoustics~\citep{brambley2008sound}, but to our knowledge have not been used in the modelling of rolling.

A description of through-thickness deformation is a prerequisite for predicting microstructural changes during rolling.
Fast rolling models, with sufficient through-thickness resolution, validated against reliable sources, remain an ongoing challenge. 
The work here develops an asymptotic model predicting inhomogeneous through-thickness deformation.
Furthermore, our results are validated against an implicit FE simulation using \Abaqus~\citep{Abaqus}. 
Our emphasis is on developing the simplest possible asymptotic model to provide a foundational understanding of the principles underlying through-thickness variation during the rolling process.  
In doing so, we neglect many factors that complicate the dynamics of rolling in reality, including anisotropic elasto-plastic material deformation in the workpiece, roll flattening, and varying friction coefficients on the rolls.   
Nonetheless, the outcomes obtained are by no means trivial, explaining the multiple pressure peaks and the seemingly complex FE observations while requiring an order of seconds to compute.
The use of small parameters in the model is justified by industrial practice. 
In a typical sheet-rolling schedule, the roll-gap thickness:length ratio typically varyies between 1:10 and 3:10, and the effective Coulomb friction coefficient is typically $0.05$--$0.2$~\citep{domanti1995two}.  
Consequently, we may assume that both the roll-gap thickness-to-length ratio and the friction coefficient are small, although due to the asymptotic framework, they are not neglected entirely.

The paper is structured as shown schematically in Figure~\ref{fig:flowchart}.
\begin{figure*}%
\centering%
\tikzset{block/.style={rectangle split, draw, rectangle split parts=2.5, rounded corners, minimum height=15em}}%
\begin{tikzpicture}[node distance = 1cm, auto]
 \node [block,rectangle split part fill={orange!20, blue!5},font=\fontsize{8}{0}\selectfont] (2Dmodel) 
 {
\hyperref[sec:governing equations]{\textbf {2D Plastic Model}}
\nodepart[text width=4cm]{two}  
\textbf{Coordinate system} 
      \begin{itemize} [wide, labelindent=8pt]
          \item Horizontal coordinate $\hat x$  
          \item Vertical coordinate $\hat y$ \\[1.5em]
      \end{itemize}
\textbf{Governing equations}
    \begin{itemize} [wide, labelindent=8pt]
        \item Mass balance
        \item Force balance  
        \item Yield criterion
        \item Flow rules  
        \item Boundary conditions  
    \end{itemize} 
      };
 \node [block,right=of 2Dmodel,node distance=7cm, rectangle split part fill={orange!20,blue!5},font=\fontsize{8}{0}\selectfont] (Multiple)  
 {
\hyperref[sec:multiplescale]{\textbf {Multiple scales}}
\nodepart[text width=4cm]{two}  
\textbf{Coordinate system} 
      \begin{itemize} [wide, labelindent=8pt]
          \item Horizontal coordinate\\[0.5em] \myquad[2] with  two distinct length \\[0.5em] \myquad[2] scales: \\[0.5em] \myquad[2]
          1- Initial half-thickness \\[0.5em] \myquad[2] $\hat h_0$
          (short length scale)\\[0.5em] \myquad[2]
          2- Length of roll gap \\[0.5em] \myquad[2] 
          $\hat l$ (long length scale)\\[0.5em] \myquad[2]
          \item Vertical coordinate $ y$ \\[1.5em] \myquad[2]
      \end{itemize}
 };
\node [block,right=of Multiple,node distance=7cm, rectangle split part fill={orange!20,blue!5},font=\fontsize{8}{0}\selectfont] (Result)  {
\hyperref[sec:result]{\textbf {Result}}
\nodepart[text width=4cm]{two}  
\textbf{Solution} 
      \begin{itemize} [wide, labelindent=8pt]
          \item Leading solution $\approx$ slab \\[0.5em] \myquad[2] model  
          \item Correction solution: fast \\[0.5em] \myquad[2] through-thickness \\[0.5em] \myquad[2] oscillations \\[1.5em]
      \end{itemize}
\textbf{Computation time}
    \begin{itemize} [wide, labelindent=8pt]
        \item Seconds  \\[3.5em] \myquad[2]  
    \end{itemize} 
      };   
\path (2Dmodel) -- node[single arrow, rounded corners=1pt, fill=orange!25, draw, align=center, xshift=0cm, yshift=-0.5cm, minimum height=2cm, minimum width=0.5cm,font=\fontsize{7}{0}\selectfont] {\hyperref[sec:non-dimensionalisation]{Scaling}} (Multiple);

\path (Multiple) -- node[single arrow, rounded corners=1pt, fill=orange!25, draw, align=center, xshift=0cm, yshift=-0.5cm, minimum height=2cm, minimum width=0.5cm,font=\fontsize{7}{0}\selectfont] {\hyperref[sec:solution]{Asymptotics}} (Result);
\end{tikzpicture}%
\caption{Outline of the development of the asymptotic multiple-scales model in section~\ref{sec:governing equations}, \ref{sec:solution}, and~\ref{sec:computing}.}%
\label{fig:flowchart}%
\end{figure*}%
The governing equations and modelling assumptions are presented in Section~\ref{sec:governing equations} where all the parameters are non-dimensionalised and the equations are scaled by two length scales using the method of multiple scales. 
The equations are then solved in Section~\ref{sec:solution} by performing asymptotic analysis on the governing equations, and the solution is provided at different asymptotic orders.
The assumption of a perfect plastic material allows for solving the stress field independent of strain, and therefore the solutions for the stress and velocity fields are provided in separate subsections~\ref{sec:solution_stress} and~\ref{sec:solution_velocity}. 
In Section~\ref{sec:computing}, the solutions are summarised and the computation methods are explained for stress and velocities in sections \ref{sec:sum_stress} and \ref{sec:sum_vel}, respectively. We have carefully constructed these sections so that readers who are not interested in the details of the multiple-scales analysis can skip Section~\ref{sec:solution} completely, and find the key outputs of the analysis and how to compute them in Section~\ref{sec:computing}. 
These outputs are validated against FE simulation results for a range of parameters in Section~\ref{sec:result}, including against FE simulations using realistic material hardening parameters in Section~\ref{sec:work-hardening}. 
The modelling assumptions made during the model derivation, and their justification from the results presented, are summarized in section~\ref{sec:assumptions}, following which implications of this work, and possibilities for future research, and discussed in Section~\ref{sec:conclusion}.

\section{Governing equations} \label{sec:governing equations}
A schematic of the rolling problem under consideration is presented in Figure~\ref{rollgap}. 
We initially work with dimensional variables, denoted with a hat, before non-dimensionalising in Section~\ref{sec:non-dimensionalisation}.  
The lateral spread in the third dimension is minimal away from the workpiece edges for sufficiently wide workpieces (width-to-thickness ratio >10~\citep{lenard2013primer}), reducing the problem to a plane strain in the two dimensions $\hat x$ and $\hat y$~\citep{domanti1995two}.  
Owing to symmetry about the sheet centre, we need only consider the upper half of the sheet. 
The rolls are held at a fixed separation that is less than the initial thickness of the sheet and the top roll rotates anti-clockwise so that the sheet moves in the positive $\hat x$-direction. 
The sheet has an initial thickness $2\hat{h}_0$ and the reduction $r$ reduces this thickness by $r\hat{h}_0$ top and bottom, giving a final thickness of $2(1-r)\hat{h}_0$. 
Attention is restricted to the region $0 \leq \hat x \leq \hat\ell$ in which the sheet is in contact with the rolls, referred to as the roll gap, where $\hat\ell = \sqrt{2r\hat{h}_0\hat{R} - r^2\hat{h}_0^2}$ is the length of the roll gap in terms of the roll radius $\hat{R}$. 
Typically, unless the reduction is very small, $\hat\ell$ is much larger than $\hat{h}_0$, giving a long thin roll gap.

The workpiece material is assumed to be perfectly plastic (i.e.\ non-hardening), and elastic effects are ignored in both the workpiece and the rolls.
This results in the same governing equations as \citet{minton2016asymptotic}.
The normal and shear Cauchy stresses, $\hat \sigma_{xx}$, $\hat \sigma_{yy}$ and $\hat \sigma_{xy}$, satisfy the von Mises yield criterion and local balance,
\begin{subequations}\label{stresses}\begin{align} 
 \frac{1}{4}\big(\hat{\sigma}_{xx}-\hat{\sigma}_{yy}\big)^2+\hat{\sigma}^2_{xy}&=\hat{\kappa}^2,\label{equ:stress1}\\
 \pdr{\hat{\sigma}_{xx}}{\hat x}+\pdr{\hat{\sigma}_{xy}}{\hat y}&=0,\\
  \pdr{\hat{\sigma}_{xy}}{\hat x}+\pdr{\hat{\sigma}_{yy}}{\hat y}&=0,
 \end{align}\end{subequations}%
where $\hat \kappa$ is the known yield stress in shear, which relates to the yield stress in tension, $\hat Y$, by $\hat Y= \sqrt{3} \hat \kappa$. 
The interface between the workpiece and the rolls is assumed to be slipping throughout the roll gap and is modelled using Coulomb friction as~$\vect{t\cdot\hat{\sigma}\cdot n} = \pm \mu \vect{n\cdot\hat{\sigma}\cdot n}$, where $\boldsymbol{n}$ and $\boldsymbol{t}$ are the unit normal and tangent vectors to the surface. 
The friction coefficient $\mu$ is also assumed to be a known constant. 
The $\pm$ sign accounts for the direction of slip of the workpiece over the rolls, as shown in Figure~\ref{rollgap}.  
Since the workpiece enters the roll gap slowly and exits it faster while the rolls rotate at a constant rate, there exists a position, $\hat x_{\text{N}}$, referred to as the neutral point, where the speed of the workpiece surface is the same as that of the rolls.  

Acquiring a better understanding of the problem's kinematics is advantageous in interpreting how the material deforms and undergoes strain, which in turn affects the microstructure. 
The unknown horizontal (rolling direction) and vertical (through-thickness direction) velocities $\hat u$ and $\hat v$, and the plastic multiplier $\hat \lambda$, satisfy mass conservation and flow rule equations, 
\begin{subequations}\label{velocities}\begin{gather} 
 \frac{\partial\hat{u}}{\partial\hat{x}}+\frac{\partial\hat{v}}{\partial\hat{y}}=0,\\
 \frac{\partial\hat{u}}{\partial\hat{x}}=\frac{1}{2}\hat{\lambda}\big(\hat{\sigma}_{xx}-\hat{\sigma}_{yy}\big),\\
 \frac{1}{2}\!\left(\frac{\partial\hat{u}}{\partial\hat{y}} + \frac{\partial\hat{v}}{\partial\hat{x}}\right)=\hat{\lambda}\hat{\sigma}_{xy}.
 \end{gather}\end{subequations}
(Note that $\hat{u}$ is horizontal velocity, and not displacement; in what follows, we always deal with velocities and never with displacements.) The velocity on the roll surfaces is restricted by the no-penetration condition $\hat V_n = 0$, where $\hat V_n$ is the normal component of sheet velocity. 
 
The boundary conditions at the entrance and exits are a given force per unit width, $\hat F_{\mathrm{in/out}}$, at each end of the roll gap.  We will also assume the shear stress distribution, $\hat \tau_0$, to be given at the entrance; we will see later (in section~\ref{sec:sum_stress}) how this may be chosen appropriately. 
Velocity boundary conditions are also applied at the entrance, where a zero vertical velocity and a constant horizontal velocity $\hat U_0$ are assumed, giving a rigid-body motion at the entrance consistent with our assumptions of symmetry, negligible elasticity, and perfect plasticity. 
Finally, the model is closed by applying symmetry about the centre line $\hat y=0$.

It is sometimes convenient to write the Cauchy stresses $\hat{\sigma}_{ij}$ in terms of the deviatoric stresses $\hat{s}_{ij}$ and the hydrostatic pressure $\hat{p}$, where, in plane strain, 
\begin{align} \label{hydrostatic}
\hat{\sigma}_{ij} &= \hat{s}_{ij}- \hat{p}\delta_{ij} &
&\text{and}&
-\hat{p}&=\frac{1}{2}(\hat{\sigma}_{xx}+\hat{\sigma}_{yy}).
\end{align}
Here, compressive stresses are negative, hence the negative sign in the definition of pressure.
We note in passing that the neglect of elasticity means the flow rule in the width direction would read $\partial\hat w/\partial\hat z = \hat\lambda\hat s_{zz}$, and so the assumption of plane strain ($\hat w=0$) implies $\hat s_{zz} = 0$ and so $\hat\sigma_{zz} = -\hat{p}$.  This is the reason for the formula for pressure in~\eqref{hydrostatic}, and the reason that $\hat\sigma_{zz}$ does not occur in the yield condition~\eqref{equ:stress1}.  A plane stress assumption would instead have $\hat\sigma_{zz}=0$ and consequently $\hat s_{zz} \neq 0$ would enter the yield condition~\eqref{equ:stress1}.

It will also be useful in the analysis below to use mass conservation applied across the whole thickness.  
Integrating the mass conservation law in~\eqref{velocities} from $\hat{y} = -\hat{h}$ to $\hat{y} = +\hat{h}$ and applying the no-penetration rule $\hat{v} = \hat{u} \mathrm{d} \hat{h}/\mathrm{d} \hat{x}$ yields
\begin{equation}\label{eq:avgmassdim}
\int_{-\hat{h}(\hat{x})}^{\hat{h}(\hat{x})} \hat{u}\ \mathrm{d} \hat{y} = 2\hat{U}_0 \hat{h}_0.
\end{equation}

\subsection{Scaling and non-dimensionalisation} \label{sec:non-dimensionalisation}

An important part of the analysis here comes from the exploitation of small dimensionless parameters. 
We therefore rescale the variables in terms of relevant dimensional scales; hats on variables denote dimensional quantities, and unhatted variables are their dimensionless equivalents.  
In what follows, distances are measured in multiples of the workpiece initial half-thickness $\hat{h}_0$, shown in Figure~\ref{rollgap}, so that the horizontal distance from the roll-gap entrance is $\hat{x} = \hat{h}_0x$ and vertical distance from the workpiece centre line is $\hat{y} = \hat{h}_0y$.  
The upper roll surface is located at $\hat{y} = \hat{h}(\hat{x}) = \hat{h}_0 h(x)$.  
Therefore, in dimensionless terms, the vertical gap between the rolls is from $y=-h(x)$ to $y=h(x)$, and the roll gap extends horizontally from $x=0$ to $x = 1/\delta$, where $\delta = \hat{h}_0/\hat{\ell}$ is the reciprocal of the roll-gap aspect ratio, and so is typically small.
All stresses are non-dimensionalised with respect to the yield stress $\hat \kappa$, and correspondingly, the entrance and exit tensions per unit width $\hat F_{\mathrm{in/out}} $ are non-dimensionalised with $\hat\kappa\hat h_0$. 
Similar to \citet{Mintonthesis}, velocities are scaled by the upstream workpiece horizontal velocity $\hat U_0 =\hat{u}(0,\hat y)$.
Finally, the plastic parameter, $\hat\lambda$, is non-dimensionalised according to its units and the preceding non-dimensionalisations with $\hat{U}_0/(\hat{h}_0 \hat{\kappa})$. 

In this study, small values of $\delta$ (corresponding to narrow roll gaps) are of interest, which is justified for cold rolling processes and is commonly assumed in many asymptotic models for cold rolling~\citep{domanti1995two, minton2016asymptotic, cherukuri1997rate}. 
Another small parameter in this model is the friction coefficient, $\mu$,
which is typically of the same order of magnitude as $\delta$.
Following previous asymptotic studies~\citep{cherukuri1997rate,cawthorn2016asymptotic,minton2016asymptotic}, we formally encoded this here by setting $\mu = \delta\beta$, where $\beta$ may be thought of as the normalised friction coefficient.

\subsubsection{Multiple scales} \label{sec:multiplescale}
The horizontal coordinate $\hat{x}$ was scaled above with a small length scale $\hat{h}_0$, yielding a rapidly-varying dimensionless horizontal variable $x$.  
While this short scale will be found later to be important, results from classical slab methods and FE analyses suggest the pressure on the roll surface slowly increases and then decreases over the length of the roll gap, $\hat\ell$, forming a profile known as the ``pressure hill''.  
This indicates a dependence of the result on large length scale $\hat\ell$ as well as on the short length scale $\hat{h}_0$. We therefore incorporate both length scales into the mathematical model using the method of multiple scales~\citep{hinch-ch7}.

We define a large-scale dimensionless horizontal coordinate $z = \delta x$ (equivalently, $z = \hat{x} / \hat{\ell}$), such that the roll-gap entrance is at $z=0$ and the exit is at $z=1$.
The shape of the rolls necessarily varies on this length scale only, and so $h(x)$ is now written $h(x) = h(z)$ to demonstrate this dependence. 
The short length scale is the sheet thickness, which is not constant. The behaviour associated with this is expected to be localised, depending only on the local properties, and therefore on the local thickness. We therefore define the short-scale horizontal coordinate, $n$, using the current thickness rather than the initial thickness by making a WKB (Wenzel--Kramers--Brillouin, sometimes WKBJ or Liouville--Green) approximation~\citep[see, e.g.][section~7.5]{hinch-ch7}
\begin{align}
\frac{\intd\hat{x}}{\intd n} &= \hat{h}(\hat{x}) &
&\Rightarrow&
n &= \int_0^{\hat x}  \frac{\intd\hat X}{\hat{h}(\hat{X})}
   = \int_0^z \frac{\intd Z}{\delta h(Z)}.
   \end{align}
In this way, $n$ measures distance through the roll gap based on the number of roll-gap-thicknesses from the entrance. We can now think of the roll gap being measured in terms of two lengths; one extending from $n=0$ to $n\sim O(1/\delta)$, and the other, from $z=0$ to $z=1$. 
If the solution $\phi(x,y)$, representing any of the mentioned variables, is assumed to depend on the short length scale through $n$ and the long length scale through $z$, then formally $\phi(x,y) = \phi(n,z,y)$, and
\begin{equation}
\pdr{\phi}{x}=\frac{1}{h}\pdr{\phi}{n}+\delta \pdr{\phi}{z}.
\end{equation}

To summarise, the non-dimensionalised horizontal distances $x$, $n$ and $z$ are defined as
\begin{align}\label{x}
x &= \hat{x}/\hat{h}_0
&
 n &= \int_0^{\hat{x}} \frac{\intd\hat X}{\hat{h}(\hat{X})}
&
z&=\hat{x}/\hat{\ell} = \delta x
\end{align}
The dependence of results on both the large and small length scales was missing in the previous asymptotic study of rolling by \citet{minton2016asymptotic}, who assumed all horizontal behaviour was only on the large length scale, and this will be shown to have led to inaccurate results.

\subsubsection{Non-dimensional governing equations} \label{sec:nondim_equs}
Based on the non-dimensionalisation and the multiple-scales variables described above, the governing equations \eqref{stresses} and \eqref{velocities} become
\begin{subequations}\label{nondim_equs}\begin{align}
 \frac{1}{4}(\sigma_{xx}-\sigma_{yy})^2+\sigma^2_{xy}&=1,
\displaybreak[0]\\
 \frac{1}{h(z)}\pdr{\sigma_{xx}}{n}+ \delta\pdr{\sigma_{xx}}{z} + \pdr{\sigma_{xy}}{y} &= 0,
\\
 \pdr{\sigma_{yy}}{y} + \frac{1}{h(z)}\pdr{\sigma_{xy}}{n} + \delta \pdr{\sigma_{xy}}{z} &= 0,
\displaybreak[0]\\
  \frac{1}{h(z)}\pdr{u}{n} + \delta \pdr{u}{ z} + \pdr{v}{y} &= 0,
\\
\frac{1}{h(z)}\pdr{u}{n}+\delta \pdr{u}{z}= \frac{1}{2}\lambda (\sigma_{xx}&-\sigma_{yy}),
\\
 \pdr{u}{y} + \frac{1}{h(z)}\pdr{v}{n} + \delta \pdr{v}{z}&=2 \lambda \sigma_{xy}.
\end{align}\end{subequations}
The Coulomb friction condition applied on the roll surface $y=h(z)$ is expressed as
\begin{multline}  \label{eq:coulomb_this}
\delta \frac{\intd h}{\intd z} \big(\sigma_{yy}-\sigma_{xx}\big) + \!\left(\!1-\delta^2\!\left(\frac{\intd h}{\intd z} \right)^{\!\!2}\right) {\sigma}_{xy}
\\
= \mp \delta \beta \!\left({\sigma}_{yy} - 2 \delta \frac{\intd h}{\intd z}  {\sigma}_{xy} + \delta^2\!\left(\frac{\intd h}{\intd z}\right)^{\!\!2}\!{\sigma}_{xx}\right),
\end{multline} 
where $\mp$ = $\sgn(x-x_{\text{N}})$ gives the correct direction of the friction force; 
here and elsewhere, we use the convention that the $-$ sign in $\mp$ refers to the zone before the neutral point ($\hat x < \hat x_{\text{N}}$), and the $+$ sign refers to the zone after the neutral point ($\hat x > \hat x_{\text{N}}$). 
Note that in~\eqref{eq:coulomb_this}, because the thickness $h(z)$ depends only on the large length scale $z$, the gradient of thickness is small, $\intd h/\intd x = \delta \intd h/\intd z$. 

Assuming non-dimensionalised tensions per unit width $F_{\mathrm{in/out}}$ are applied at the entrance and exit,
\begin{equation} F_{\mathrm{in/out}}= \int_{-h_{\mathrm{in/out}}}^{h_{\mathrm{in/out}}} \sigma_{xx} \, \intd y, \end{equation}
where $h_{\mathrm{in}} = 1 $ by our non-dimensionalization, and $h_{\text{out}}$ is half of the final thickness imposed by the rolls. 
Velocity and shear stress are also assumed to be prescribed at the entrance as,
\begin{align} \label{eq:vel_ic}
    u(z=n=0) &= 1, &
    v(z=n=0) &= 0,&
    &\text{and}&
    \sigma_{xy} &= \delta \tau_0,
\end{align}
where we assume that the imposed shear stress at the entrance is $O(\delta)$, and so $\tau_0 = \hat \tau_0 /(\delta\hat \kappa) $ is the non-dimen\-sion\-alisation used.  This will turn out below to be the correct number of boundary conditions to apply, as will be seen once we have derived the solution.

The no-flux constraint on the roll surface $y= h(z)$ and the averaged mass balance~\eqref{eq:avgmassdim} become
\begin{align}\label{eq:avgmassnd}
v &= \delta \frac{\intd h}{\intd z}u,
&&\text{and}&
\int_{-h(z)}^{h(z)} u\ \mathrm{d} y &= 2.
\end{align}
Finally, symmetry about the centre line at $y=0$ is applied,
\begin{align} \label{eq:midplane}
   {\sigma}_{xy} ( y=0)&=0 &&\text{and}& v( y=0)&=0.
\end{align}

Our next step is to take the model written in terms of the two longitudinal length scales \eqref{nondim_equs}-\eqref{eq:midplane}, and analyse it using asymptotic analysis.  
This procedure is outlined in Section \ref{sec:solution} below. 
We note that the mathematical description of the rolling system does temporarily become more complicated as a result of the multiple length scales and asymptotic analysis, but upon completion yields a simpler and faster-to-compute model. Readers uninterested in the technical details may safely skip to Section \ref{sec:computing} where the outputs of the model are summarised before being compared to FE simulations in Section~\ref{sec:result}.


\section{Asymptotic solution} \label{sec:solution}

This section solves \eqref{nondim_equs}-\eqref{eq:midplane} using an asymptotic expansion in powers of the small parameter $\delta$.  
The results are summarised at the beginning of section~\ref{sec:computing}.  
In order to solve equations \eqref{nondim_equs}-\eqref{eq:midplane}, the stress components, velocity components, and plastic parameter
are expanded as asymptotic series in the small parameter $\delta$,
\begin{equation}\label{expansion} \phi= \phi^{(0)}(z,y)+ \delta \phi^{(1)}(n,z,y) + \delta^2 \phi^{(2)}(n,z,y) + O(\delta ^3), 
\end{equation}
where $\phi$ represents any of the mentioned variables.  
Under the assumption that $\delta$ is small, each power of $\delta$ represents a small correction to the terms preceding it. 
Note that the leading-order terms in all variables are set to be independent of $n$, which is in line with the slab method and will be justified a posteriori. 

We now proceed to solve the problem presented in Section \ref{sec:nondim_equs} by substituting the expansion \eqref{expansion} into the non-dimensionalised governing equations~\eqref{nondim_equs} and boundary conditions \eqref{eq:coulomb_this} to \eqref{eq:midplane} and collecting terms with like powers of $\delta$. 
Anticipating that the stress is independent of velocity, we first derive a solution for the stress components, before returning to the velocities in Section~\ref{sec:solution_velocity}.

\subsection{Solving for the stresses} \label{sec:solution_stress}
 By expanding in powers of $\delta$ and collecting similar terms, equations~\eqref{nondim_equs} become
\begin{subequations}\begin{align}
\label{eq:yield}
&\frac{1}{4}\Big(\sigma_{xx}^{(0)} -\sigma_{yy}^{(0)} \Big)^{\!2} + \sigma_{xy}^{(0)}\strut^2
\\\notag&\qquad
+ \delta \!\left(\frac{1}{2}(\sigma_{xx}^{(0)} -\sigma_{yy}^{(0)})(\sigma_{xx}^{(1)} -\sigma_{yy}^{(1)})  + 2\sigma_{xy}^{(0)}\sigma_{xy}^{(1)}\!\right)\!
\\\notag&\qquad
+ \delta^2 \!\left(\frac{1}{4}\!\left[(\sigma_{xx}^{(1)} -\sigma_{yy}^{(1)})^2 + 2(\sigma_{xx}^{(0)} -\sigma_{yy}^{(0)})(\sigma_{xx}^{(2)} -\sigma_{yy}^{(2)})\right]
\right.\\\notag&\qquad\qquad\quad\left.
{}+ \sigma_{xy}^{(1)}\strut^2 + 2 \sigma_{xy}^{(0)}\sigma_{xy}^{(2)}\!\right)\! + O(\delta ^3) = 1,
\displaybreak[0]\\
\label{eq:mom1}
&\pdr{\sigma_{xy}^{(0)}}{y} + \delta \!\left(\!  \pdr{\sigma_{xx}^{(0)}}{z}+ \frac{1}{h(z)}\pdr{\sigma_{xx}^{(1)}}{n} +  \pdr{\sigma_{xy}^{(1)}}{y} \!\right)\!
\\\notag&\qquad
+ \delta^2 \!\left(\! \pdr{\sigma_{xx}^{(1)}}{z}+ \frac{1}{h(z)}\pdr{\sigma_{xx}^{(2)}}{n} +  \pdr{\sigma_{xy}^{(2)}}{y} \!\right)\! + O(\delta ^3)= 0, \displaybreak[0]
\displaybreak[0]\\
\label{eq:mom2}
&\pdr{\sigma_{yy}^{(0)}}{y}  + \delta \!\left(\! \pdr{\sigma_{yy}^{(1)}}{y} +\pdr{\sigma_{xy}^{(0)}}{z} + \frac{1}{h(z)}\pdr{\sigma_{xy}^{(1)}}{n} \!\right)\!
\\\notag&\qquad
+ \delta^2 \!\left(\! \pdr{\sigma_{yy}^{(2)}}{y}+ \pdr{\sigma_{xy}^{(1)}}{z} + \frac{1}{h(z)}\pdr{\sigma_{xy}^{(2)}}{n} \!\right)\! + O(\delta ^3)= 0. 
\end{align} \end{subequations}%
The Coulomb friction surface boundary conditions for stress from~\eqref{eq:coulomb_this} becomes
\begin{align} \label{eq:friction}  
&\sigma_{xy}^{(0)} + \delta \left( \frac{\intd h}{\intd z} (\sigma_{yy}^{(0)}-\sigma_{xx}^{(0)}) + \sigma_{xy}^{(1)} \right)
\\&\qquad\notag
+ \delta^2 \left( \frac{\intd h}{\intd z} (\sigma_{yy}^{(1)}-\sigma_{xx}^{(1)}) + \sigma_{xy}^{(2)}  - \!\left(\frac{\intd h}{\intd z}\right)^{\!\!2}\! \sigma_{xy}^{(0)} \right) + O(\delta ^3)  
\\&\quad\notag
= \mp \delta \left(  \beta \sigma_{yy}^{(0)} \right) \mp \delta^2 \left(  \beta \sigma_{yy}^{(1)} - 2 \beta \frac{\intd h}{\intd z} \sigma_{xy}^{(0)} \right) + O(\delta ^3).
\end{align}
Forward/backward tension per unit width at the roll-gap entrance/exit can be expanded as 
\begin{equation}\label{eq:boundary tension} F_{\mathrm{in/out}}= \int_{-h_{\mathrm{in/out}}}^{h_{\mathrm{in/out}}} \sigma^{(0)}_{xx}  \, \intd y  + \delta \int_{-h_{\mathrm{in/out}}}^{h_{\mathrm{in/out}}} \sigma^{(1)}_{xx} \, \intd y + O(\delta ^2). \end{equation}
Shear stress is assumed to have a known distribution at the entrance, $\delta \tau_0$, and hence
\begin{equation} \label{eq:shear_ic}
     \sigma_{xy}(z=n=0) = 0 + \delta \tau_0 + O(\delta ^2),
\end{equation}
Finally, symmetry condition implies that ${\sigma}_{xy}$ is zero at all orders of $\delta$ about the centre line at $y=0$.

\subsubsection{Leading-order solution} \label{sec:leadingstress}

At leading order, the friction equation~\eqref{eq:friction} implies $\sigma_{xy} ^ {(0)} = 0$ on the roll surface, and the local balance equation~\eqref{eq:mom1} implies that it is independent of $y$.  
Hence, $\sigma_{xy}^{(0)}=0$ everywhere; i.e.\ the shear stress $\sigma_{xy}$ is small (at most $O(\delta)$) throughout the roll gap.  
At leading order, the yield equation~\eqref{eq:yield} gives 
\begin{gather}
\frac{1}{4}\Big(\sigma_{xx}^{(0)} -\sigma_{yy}^{(0)} \Big)^2=1
\\\notag\begin{aligned}
&\Rightarrow&
\sigma_{xx}^{(0)}&= 1-p^{(0)}
&&\text{and}& 
\sigma_{yy}^{(0)}&= -1-p^{(0)},
\end{aligned}\end{gather}
where the leading-order hydrostatic pressure $-2p^{(0)} = \sigma_{xx}^{(0)} +\sigma_{yy}^{(0)}$, and we have asserted that $\sigma_{yy}$ is more compressive than $\sigma_{xx}$. 
From local balance equation~\eqref{eq:mom2} at leading order, $\sigma_{yy}^{(0)}$ is independent of $y$, and so consequently are $p^{(0)}$ and $\sigma_{xx}^{(0)}$. 
This implies that both normal stress components are vertically homogeneous at this order.
The solution for $p^{(0)}(z)$ will be dictated by satisfying the friction expression \eqref{eq:friction} at $O(\delta)$ on the surface, although this involves the as-yet-unknown $\sigma_{xy}^{(1)}$, and so we must first solve the first-order equations before $p^{(0)}$ can be fully determined.  
Boundary conditions for $p^{(0)}$ are determined by the prescribed front and back tension at the entrance and exit \eqref{eq:boundary tension},
\begin{align} \label{eq:p_zero boundary} p^{(0)}(z=0)&=1 - \frac{F_{\mathrm{in}}}{2}, &
 p^{(0)}(z=1)&=1 - \frac{F_{\mathrm{out}}}{2h(1)}.
\end{align}

 \subsubsection{First-order solution} \label{sec:firststress}
 By taking the terms of order $\delta$ in the yield function \eqref{eq:yield}, and from the definition of hydrostatic pressure, $\sigma_{xx}^{(1)}$ and $\sigma_{yy}^{(1)}$ are given by 
\begin{equation} \label{eq:sigma_first} \sigma_{xx}^{(1)}(n,y,z) =\sigma_{yy}^{(1)}(n,y,z)  = -p^{(1)}(n,y,z) . \end{equation}
The local balance equations \eqref{eq:mom1} and \eqref{eq:mom2} at this order then become
\begin{subequations}\label{eq:mom_first}\begin{align}
\pdr{\sigma_{xy}^{(1)}}{y} -\frac{1}{h(z)}\pdr{p^{(1)}}{n} &= \frac{\intd p^{(0)}}{\intd z},\label{eq:mom_first_a}\\
 \pdr{p^{(1)}}{y} - \frac{1}{h(z)} \pdr{\sigma_{xy}^{(1)}}{n} &=0.\label{eq:mom_first_b}
\end{align}\end{subequations}
These two first-order equations combined to give a single second-order wave equation for either $p^{(1)}$ or $\sigma_{xy}^{(1)}$, with $n$ playing the role of the temporal variable and $y$ the role of the spacial variable. For example, taking $\partial(\ref{eq:mom_first_b})/\partial y + (1/h)\partial(\ref{eq:mom_first_a})/\partial n$ gives a wave equation for $p^{(1)}$.  The general solution to~\eqref{eq:mom_first} is therefore found to be a wave solution given by
\begin{subequations}\begin{align}
p^{(1)} &= A_1\!\left(n+\frac{y}{h},z\right)+A_2\!\left(n-\frac{y}{h},z\right)+D(z),\\
\sigma_{xy}^{(1)} &= A_1\!\left(n+\frac{y}{h},z\right)-A_2\!\left(n-\frac{y}{h},z\right)+y\frac{\intd p^{(0)}}{\intd z}.
\end{align}\label{eq:p_shear_first}\end{subequations}%
The function $A_1(\xi,z)$ and $A_2(\xi,z)$ are the as-yet-unknown wave forms moving in the negative and positive directions respectively.  However, by symmetry about $y=0$ we know that $p$ is symmetric in $y$ and $\sigma_{xy}$ is antisymmetric in $y$, and hence we must have that $A_1(\xi,z) = A_2(\xi,z) = A(\xi,z)$ for some as-yet-unknown wave form $A(\xi,z)$.  The function $D(z)$ is a constant of integration  with respect to $n$ and $y$. 
Equations \eqref{eq:p_shear_first} are travelling waves in the fast axial variable $n$, and hence result in rapid stress oscillations. 
This finding distinguishes this study from previous asymptotic models, where all horizontal behaviour was assumed to be at the long length scale, and consequently such oscillatory patterns were not captured. 

With the general form of the solution for $\sigma_{xy}^{(1)}$ now known, we may now solve for $p^{(0)}$, which remained undetermined at leading order.  
At $O(\delta)$, the Coulomb friction equation~\eqref{eq:friction} is
\begin{equation} \label{eq:coulomb-first}
    -2\frac{\intd h}{\intd z} + \sigma_{xy}^{(1)} \mp \beta (1+p^{(0)}) = 0. 
\end{equation}
Substituting $\sigma_{xy}^{(1)}$ from~\eqref{eq:p_shear_first} into~\eqref{eq:coulomb-first} and evaluating at $y=h$ results in
 \begin{equation}  \label{eq:coulomb-first_ext} 
 -2\frac{\intd h}{\intd z} \mp \beta (1+p^{(0)}) + h\frac{\intd p^{(0)}}{\intd z} = - \big(A(n+1,z)-A(n-1,z)\big),\end{equation}
where the left-hand side is a function of the slow variable $z$ only and is independent of the fast variable $n$.
If the left hand side is nonzero, then $A$ grows by this much every time $n$ increases by $2$, and by the end of the roll gap, $p^{(1)}$ or $\sigma_{xy}^{(1)}$ would have grown to be $O(n) = O(1/\delta)$ and the asymptotic ordering we assumed in deriving our equations would be broken~\citep{hinch-ch7}. We therefore require~\eqref{eq:coulomb-first_ext} to be zero, and so we take
\begin{align}\label{eq:A:global}
    A(n+1,z) - A(n-1,z) &= 0,
    &&\text{and}&
     \int ^ 1 _ {-1}  A(\xi,z)\,\intd\xi &= 0;
\end{align}
the latter condition achieved by shifting $A$ by a function of $z$ which is then absorbed into $D$.
Substituting~\eqref{eq:A:global} into~\eqref{eq:coulomb-first_ext} gives a first-order ordinary differential equation (ODE) for $p^{(0)}(z)$ as,
\begin{equation} \label{eq:p_lead}
     -2\frac{\intd h}{\intd z} \mp \beta (1+p^{(0)}) + h\frac{\intd p^{(0)}}{\intd z} = 0.
\end{equation}
This is analogous to the ODE for $\sigma_{yy}^{(0)}$ derived previously by \citet{minton2016asymptotic} for asymmetric thin sheet rolling and by \citet{cawthorn2016asymptotic} for sandwich rolling. 
This can be solved with the boundary conditions \eqref{eq:p_zero boundary} once from the entrance forwards with $-$ve sign of the coefficient of the friction term (i.e.\ assuming slip is due to the rolls moving faster than the sheet) and once from the exit backwards with $+$ sign (i.e.\ assuming slip is due to the sheet moving faster than the rolls). 
Where these two solutions cross determines the location of the neutral point $z_{\text{N}}$, and the whole solution gives the expected ``pressure hill'' obtained from classical slab methods. 
In fact, the approximation obtained here by the asymptotic analysis at leading order (by restricting the effect of shear stress to a small contribution on the surface) is identical to that obtained by the slab method.
The velocities will be confirmed to give slipping in the correct direction either side of the neutral point $z_\mathrm{N}$ in section~\ref{sec:solution_velocity} below.

Returning to the first-order solution presented in \eqref{eq:p_shear_first}, we now consider the two unknown functions $A(\xi,z)$ and $D(z)$. 
Condition~\eqref{eq:A:global} establishes that, as a function of $\xi$, $A$ is periodic with period 2. 
This means $A$ needs only be found for $-1 < \xi <1$ as $z$ varies to be fully determined.  
Finding the $z$-dependency of the function $A$, as well as the unknown $D(z)$, requires further information which will be revealed in the next order of correction. 
However, the boundary condition at the entrance provides the initial conditions. 
Since $F_{\mathrm{in}}$ has already been satisfied at leading order, the first-order correction to it must be zero. 
Hence, according to equation \eqref{eq:boundary tension} and using condition \eqref{eq:A:global},
\begin{align} \label{eq:D_ic}
0 &= \int_{-h_0} ^{h_0} \sigma^{(1)}_{xx} \left(y,\,0\right)\,\intd y
\\\notag
&= - \int_{-1}^{1} \big(A\!\left(y,\,0\right)\!+A\left(-y,\,0\right)\!-D(0)\big) \intd y
= 2D(0),
\end{align}
which is an initial condition for $D(z)$. Unlike the leading order, where the average force was sufficient for the stress boundary conditions, at this order, the force distribution must be specified. We therefore assume a known initial stress distribution at the entrance, given by $\sigma^{(1)}_{xx}(y,0)$ and $\tau_0(y)$. 
At $z=0$, \eqref{eq:sigma_first} and~\eqref{eq:p_shear_first} then give
\begin{subequations}
    \begin{gather}
        \sigma_{xx}^{(1)}(y,0) = -A(y,0) - A(-y,0), \\
        \tau_{0}(y) = A(y,0) - A(-y,0) + y \frac{\intd p^{(0)}}{\intd z}.
    \end{gather}
\end{subequations}
and consequently,
\begin{equation} \label{eq:A_ic}
A \left(y,\,0\right) = \frac{1}{2}\!\left( \tau_{0}(y) - \sigma_{xx}^{(1)}(y,0) \right) -\frac{1}{2} y\!\left.\frac{\intd p^{(0)}}{\intd z}\right|_{\mathrlap{z=0}} \quad,
\end{equation}
Equation~\eqref{eq:A_ic} gives an initial condition for $A$ at $z=0$.  The choice of initial stresses will be discussed later in Section~\ref{sec:sum_stress}.

\subsubsection{Second-order solution} \label{sec:secondstress}

The solution is continued to this order with the goal of finding the unknown parameters $A(\xi,z)$ and $D(z)$ at the previous order. 
From the yield function \eqref{eq:yield} at $O(\delta^2)$, we find that
\begin{align}
&\sigma_{xx}^{(2)} = -p^{(2)} - \frac{1}{2} \sigma_{xy}^{(1)} \strut^2  &\text{and}&  &\sigma_{yy}^{(2)} = -p^{(2)} + \frac{1}{2} \sigma_{xy}^{(1)} \strut^2.
\end{align}
Using the local balance equations~\eqref{eq:mom1} and~\eqref{eq:mom2} at $O(\delta^2)$, and substituting the known solution for $\sigma_{xx}^{(1)}$ from~\eqref{eq:sigma_first}, we find a wave equation for $\sigma_{xy}^{(2)}$ and $p^{(2)}$ analogous to the first order wave equation~\eqref{eq:mom_first},
\begin{subequations}\label{eq:wave_second}\begin{align}
  \pdr{\sigma_{xy}^{(2)}}{y} -  \frac{1}{h}\pdr{p^{(2)}}{n}&= \pdr{p^{(1)}}{z} + \frac{1}{2h}\pdr{\sigma_{xy}^{(1)} \strut^2}{n} &
\\
 \pdr{p^{(2)}}{y} - \frac{1}{h}\pdr{\sigma_{xy}^{(2)}}{n}&= \pdr{\sigma_{xy}^{(1)}}{z} + \frac{1}{2}\pdr{\sigma_{xy}^{(1)} \strut^2}{y} . 
\end{align}\end{subequations}%
Substituting  $p^{(1)}$ and  $\sigma_{xy}^{(1)}$ from equation \eqref{eq:p_shear_first} into~\eqref{eq:wave_second}, solving for $\sigma_{xy}^{(2)}$ and $p^{(2)}$, and substituting this solution into the friction equation~\eqref{eq:friction} at $O(\delta^2)$, leads eventually after significant but uninteresting algebra (detailed in appendix~\ref{App:sigma_xy_second}) to
\begin{align} \label{eq:A}
  &2h\pdr{A(n+1,z)}{z}
  \\\notag&\qquad
  + \pdr{}{n} \Big(\! A(n+1,z)^2 \Big)
  +\left(\!h  \frac{dp^{(0)}}{dz} \mp 2 \beta\!\right)\!A(n+1,z) 
  \\\notag&\qquad
   - h\frac{\intd D}{\intd z} \mp \beta D(z)
   = -\big[M(n+1,z) - M(n-1,z)\big],
\end{align}
where $M(\xi,z)$ is the second order wave-equation homogeneous solution to~\eqref{eq:wave_second} analogous to the first order wave-equation solution $A(\xi,z)$.  We do not need to calculate $M$, as by the same secularity argument used for $A$ at first order, we need only require that $M$ be bounded in $n$, and hence that the right-hand side of equation \eqref{eq:A} be zero.
Integrating~\eqref{eq:A} between $n=-2$ and $n=0$ and imposing~\eqref{eq:A:global} then gives
\begin{equation}
     -h\frac{\intd D}{\intd z} \mp \beta D(z) =0.
\end{equation} 
Since $D(0)=0$ from~\eqref{eq:D_ic}, we conclude that $D(z) \equiv 0$. 
With this, equation~\eqref{eq:A} then gives an evolution equation for how $A$ varies as $z$ is increased,
\begin{equation} \label{eq:simplified_sigma}
  2h\pdr{A(\xi,z)}{z}   + \pdr{}{\xi} \Big(\! A(\xi,z) \strut ^ 2 \Big) + \!\left(\! h  \frac{\intd p^{(0)}}{\intd z} \mp 2 \beta \!\right)\!A(\xi,z) = 0. \end{equation} 
This equation may be rewritten as a standard Burger's equation by making the substitution
\begin{subequations}\label{eq:Burgers_total}\begin{gather}
T  = \int^z \frac{\alpha_1}{h(\bar z)^2}\ \mathrm{d} \bar z,
\qquad\qquad
\omega\big(\xi,T(z)\big) = \frac{A(\xi,z)}{(\alpha_1 / h)},
\\
\alpha_1 = \exp \Bigg\{\! \int ^ z _0 \mp \frac{\beta(p^{(0)}(\Tilde{z}) - 1)}{2h(\Tilde{z})} \intd\Tilde{z} \Bigg\},
\\  
\Rightarrow\qquad
\pdr{\omega}{T} + \frac{1}{2}  \pdr{}{\xi} \Big( \omega^ 2 \Big) = 0.
\label{eq:Burger}\end{gather}\end{subequations}
This is especially advantageous as it enables the effective handling of the evolution of discontinuities in the form of shocks or expansion fans.
Once~\eqref{eq:Burger} has been solved with suitable initial conditions for $\omega$ and hence $A$, the correction to the pressure, $p^{(1)}$, and all stress components $\sigma_{xx} ^{(1)}$, $\sigma_{yy}^{(1)}$ and $\sigma_{xy}^{(1)}$, can be calculated by evaluating \eqref{eq:p_shear_first} and \eqref{eq:sigma_first}. 
Since the boundary condition changes on the surface at the neutral point, equation \eqref{eq:Burger} should be solved
separately for the entrance and exit regions. 
The behaviour at their intersection in the vicinity of the neutral point is discussed further in the next section.


\subsubsection{Behaviour near the neutral point}\label{sec:neutral_point}
 
\begin{figure*}%
\centering%
\includegraphics{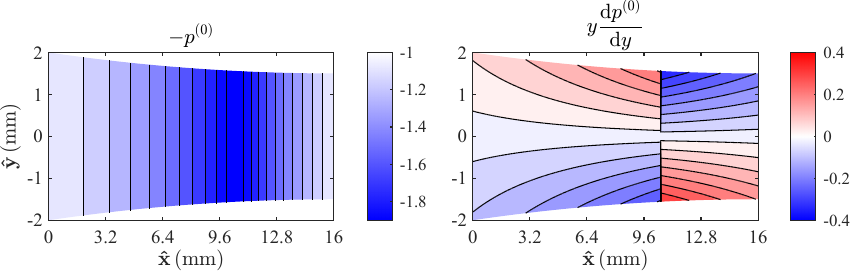}\par%
\caption{Leading-order pressure (left) and the contribution to shear $y\,\intd p^{(0)}\!/\intd z$ (right). Results are scaled with $\hat{\kappa}$, the yield stress in shear. Parameters used are $(\hat h_0, \delta, r, \mu)=(2\,\mathrm{mm}, 0.125, 25\%, 0.1)$.}%
\label{fig:slab} %
\end{figure*}

The shear stress $\sigma_{xy} ^{(1)}$, given in equation~\eqref{eq:p_shear_first}, is formed of two components: the first component involving $A$ will be seen to cause a wave pattern, while the second, $y\,\intd p^{(0)}\!/\intd z$,
changes suddenly at the neutral point.  This could hypothetically cause a discontinuity in shear stress across a straight line running vertically through the thickness of the sheet, as illustrated in the right panel of Figure~\ref{fig:slab}. 
This discontinuity is predicted by many mathematical models for cold rolling that use Coulomb friction \citep{smelserjohnson1992asymptotic,domanti1995two,cherukuri1997rate,minton2016asymptotic}.  
Physically, the sudden change in sign of the shear stress at the surface is an expected consequence of Coulomb friction, but any through-thickness discontinuity must be artificial, since otherwise vertical forces would not balance on the thin vertical slice of material at the neutral point. 
Instead, we require $\sigma_{xy}$ to be continuous in $z$.
Using the notation $(-)$ and $(+)$ to denote variables to the left and right of the neutral point, near the neutral point $z=z_{\text{N}}$ we require $\sigma_{xy}^{(1-)} = \sigma_{xy}^{(1+)}$. Therefore from~\eqref{eq:p_shear_first} we obtain,
\begin{align} \label{eq:xy_inner}
&A^{(-)}\!\left(n+\frac{y}{h},z\right)-A^{(-)}\!\left(n-\frac{y}{h},z\right)
\\ &\notag\qquad\qquad
+\frac{h_{\text N}}{2} \!\left[ \left(n+\frac{y}{h}\right) - \left(n-\frac{y}{h}\right) \right]\! \frac{\intd p^{(0-)}}{\intd z} \Big| _{z=z_{\text{N}}}
\\ &\notag\quad
= A^{(+)}\!\left(n+\frac{y}{h},z\right)-A^{(+)}\!\left(n-\frac{y}{h},z\right)
\\ &\notag\qquad\qquad
+\frac{h_{\text N}}{2} \!\left[ \left(n+\frac{y}{h}\right) - \left(n-\frac{y}{h}\right) \right]\! \frac{\intd p^{(0+)}}{\intd z} \Big| _{z=z_{\text{N}}}.
 \end{align} %
Solving \eqref{eq:xy_inner} and applying~\eqref{eq:p_lead} yields
\begin{align} \label{eq:Adiscontinuity}
    A^{(+)}(\xi,z_{\text{N}})
    &= A^{(-)}(\xi,z_{\text{N}}) + \frac{h_{\text{N}}}{2} \xi\!\left(\!\frac{\intd p^{(0-)}}{\intd z} \Big| _{z_{\text{N}}}\!\! - \frac{\intd p^{(0+)}}{\intd z} \Big| _{z_{\text{N}}} \right)
    \notag\\&
    = A^{(-)}(\xi,z_{\text{N}}) + \beta \xi \!\left(1+ p^{(0)} (z_{\text{N}}) \!\right)\!.
\end{align}
In this derivation, we assumed that there is no $O(\delta)$ correction to the location of the neutral point, and that the change from $A^{(-)}$ to $A^{(+)}$ happens abruptly. 
These assumptions are verified via a more thorough calculation about the neutral point in Appendix~\ref{appendix:neutral_point}. Since both $A^{(-)}$ and $A^{(+)}$ are periodic in $\xi$ with period 2, equations~\eqref{eq:Adiscontinuity} completely define $A^{(+)}(\xi,z_{\text{N}})$ in terms of $A^{(-)}$, from which equation~\eqref{eq:simplified_sigma} or~\eqref{eq:Burger} can be used to evolve $A^{(+)}$ for $z_{\text{N}}<z<1$. 
Note that $A^{(-)}$ is known completely from the initial conditions~\eqref{eq:A_ic} at $z=0$ and then solving equation~\eqref{eq:simplified_sigma} or~\eqref{eq:Burger} for $0<z<z_{\text{N}}$.  

\subsection{Solving for the velocities} \label{sec:solution_velocity}
Having solved for the stresses in Section \ref{sec:solution_stress}, we now begin the calculation of the velocity field. 
The approach here is the same as was used for the stresses above. 
By asymptotically expanding equations~\eqref{nondim_equs} in successive powers of $\delta$ and collecting like terms, we obtain
\begin{subequations}\label{eq:vel_expand}%
\begin{align}\label{eq:continuity} 
&\pdr{v^{(0)}}{y} + \delta \!\left(\! \pdr{u^{(0)}}{z} +  \frac{1}{h(z)}\pdr{u^{(1)}}{n}+ \pdr{v^{(1)}}{y}  \!\right)\!
\\\notag&\qquad
+ \delta ^2  \!\left(\! \pdr{u^{(1)}}{z} + \frac{1}{h(z)}\pdr{u^{(2)}}{n} + \pdr{v^{(2)}}{y} \!\right)\! + O(\delta ^3)= 0,
\displaybreak[0]\\
&\delta \!\left(\! \pdr{u^{(0)}}{z} +  \frac{1}{h(z)}\pdr{u^{(1)}}{n}  \!\right)\! + \delta^2 \!\left(\! \pdr{u^{(1)}}{z} +  \frac{1}{h(z)}\pdr{u^{(2)}}{n}  \!\right)\!  + O(\delta ^3)
\notag\\&\qquad\label{eq:flowtension} 
= \frac{1}{2}  \lambda^{(0)}  \!\left(\!\sigma_{xx}^{(0)}  - \sigma_{yy}^{(0)} \!\right)
\\\notag&\qquad\quad
+ \frac{\delta}{2} \!\left(\!\lambda^{(0)}  \!\left(\!\sigma_{xx}^{(1)}  - \sigma_{yy}^{(1)} \!\right)\!
+  \lambda^{(1)}  \!\left(\!\sigma_{xx}^{(0)}  - \sigma_{yy}^{(0)} \!\right) \!\right)\!+ O(\delta ^2),\displaybreak[0]\\&%
\label{eq:flowshear} 
\pdr{u^{(0)}}{y}  + \delta \!\left(\! \pdr{u^{(1)}}{y}  + \pdr{v^{(0)}}{z} + \frac{1}{h(z)}\pdr{v^{(1)}}{n} \!\right)\!
\\\notag&\qquad\quad
+ \delta^2 \!\left(\! \pdr{u^{(2)}}{y} +  \pdr{v^{(1)}}{z} +  \frac{1}{h(z)}\pdr{v^{(2)}}{n} \!\right)\! + O(\delta ^3)
\\\notag&\qquad
= 2 \sigma_{xy}^{(0)} \lambda^ {(0)} + 2\delta \!\left(\! \lambda^ {(0)} \sigma_{xy}^{(1)} + \lambda^ {(1)} \sigma_{xy}^{(0)} \!\right)\! 
\\\notag&\qquad\quad
+ 2\delta^2 \!\left(\! \lambda^ {(0)}  \sigma_{xy} ^{(2)} +\lambda^ {(1)}  \sigma_{xy} ^{(1)}+\lambda^ {(2)}  \sigma_{xy} ^{(0)} \!\right)\! + O(\delta ^3).%
\end{align}\end{subequations}

The surface boundary condition for velocity from~\eqref{eq:avgmassnd} is,
\begin{equation}
\label{eq:boundary2 velocity}
v^{(0)} + \delta v^{(1)} + \delta^ 2 v^{(2)} + O(\delta ^3) = \delta \frac{\intd h}{\intd z} \left( u^{(0)} + \delta u^{(1)} \right) + O(\delta ^3). 
\end{equation}

For the velocity initial condition, at $z=0$, we set $v^{(1)} \equiv u^{(1)} \equiv 0$ for consistency with our assumptions. This reflects that the sheet has undergone negligible elastic deformation at the entrance, and there is no plastic deformation outside the roll gap, so that the sheet enters the roll gap as a solid body motion.
Thus, equation \eqref{eq:vel_ic} is expanded as,
\begin{subequations}\label{eq:vel_ic_expand}\begin{align} 
    u(z=n=0) &= 1 + \delta (0) + \delta ^2 (0) + O(\delta ^3)
\\
    v(z=n=0)&= 0 + \delta (0) + \delta ^2 (0) + O(\delta ^3)
\end{align}\end{subequations}%
Finally, the symmetry condition \eqref{eq:midplane} implies $v$ is zero at all orders of $\delta$ about the centre line at $y=0$.

\subsubsection{Leading-order}
Evaluating \eqref{eq:flowshear} at leading order yields $\partial u^{(0)}/\partial y = 0$,
indicating $u^{(0)}$ is independent of $y$, as for the leading order stresses. 
Conservation of mass~\eqref{eq:avgmassnd} at leading order then gives
\begin{equation}
u^{(0)}(z)=\frac{1}{h(z)}.
\label{eq:u0}
\end{equation}
Note that the velocity of the rolls has not been required in our analysis. Given that~\eqref{eq:u0} shows the material velocity increases monotonically, and that in solving for the neutral point $z_N$ above we assumed the slip over the rolls changed direction at the neutral point, we can infer the velocity of the rolls to be $U_\mathrm{rolls} = u^{(0)}(z_N) + O(\delta)$.  This is consistent with our non-dimensionalisation above, where velocities were non-dimensionalised based on the sheet entrance velocity $\hat{U}_0$, and therefore the speed of the rolls was unknown.

The vertical velocity at leading order, $v^{(0)}$, is identically zero. 
This can be determined by evaluating the continuity equation~\eqref{eq:continuity} at leading order to show that $v^{(0)}$ is independent of $y$, and using the no-penetration boundary condition~\eqref{eq:boundary2 velocity}, which imposes zero vertical velocity $v^{(0)}$ on the surface. 
Based on the tension flow rule equation~\eqref{eq:flowtension}, the leading-order plastic multiplier $\lambda^{(0)}$ is also found to be zero.
\subsubsection{First-order}
At $O(\delta)$, the continuity equation \eqref{eq:continuity} and the shear flow-rule \eqref{eq:flowshear} become
\begin{align}
    &\pdr{v^{(1)}}{y}+\frac{1}{h}\pdr{u^{(1)}}{n}=-\pdr{u^{(0)}}{z}&
    &\text{and}&
    &\pdr{u^{(1)}}{y}+\frac{1}{h}\pdr{v^{(1)}}{n}=0 .
\end{align}
This is a wave equation for $u^{(1)}$ and $v^{(1)}$, similar to the stresses at this order, with solution
\begin{subequations}\label{eq:vel_first}\begin{align}
u^{(1)} &= B\!\left(n+\frac{y}{h},z\right) +B\!\left(n-\frac{y}{h},z\right),  \\
v^{(1)} &= -\left[B\!\left(n+\frac{y}{h},z\right) -B\!\left(n-\frac{y}{h},z\right) \right] + \frac{y}{h^2} \frac{\intd h}{\intd z},
\end{align}\end{subequations}%
where $B(\xi,z)$ is an unknown function, and we have assumed symmetry of $u$ and asymmetry of $v$. 
By substituting the solution for $v^{(1)}$ into the tension flow-rule~\eqref{eq:flowtension}, $\lambda^{(1)}$ is found to be
\begin{align}  \label{eq:lambda_first} \lambda^{(1)} &= -\pdr{v^{(1)}}{y} 
\\\notag&
= \frac{1}{h}\left[B'\!\left(n+\frac{y}{h},z\right) +B'\!\left(n-\frac{y}{h},z\right) \right] -  \frac{1}{h^2} \frac{\intd h}{\intd z}, \end{align}
with prime denoting $\partial/\partial \xi$.
No-penetration boundary condition on the surface \eqref{eq:boundary2 velocity} then gives
 \begin{align} \label{eq:Bperiodic}
 &v^{(1)}(y=h) = \frac{\intd h}{\intd z} u^{(0)}  
 &\Rightarrow& &B(n+1,z)=B(n-1,z),\end{align}
which implies that $B(\xi,z)$ is periodic with periodicity~2, similarly to $A(\xi,z)$ for the stresses; hence, 
$B$ need only be defined for $-1<\xi<1$.  
From~\eqref{eq:vel_ic_expand}, by imposing zero first-order velocities at the entrance, the initial conditions for $B(\xi,z)$ at $n=z=0$ are given from~\eqref{eq:vel_first} as
\begin{equation}
B(y,\,0) =\frac{y}{2}\left.\frac{\intd h}{\intd z}\right|_{z=0}.
\label{eq:v_ic}
\end{equation}
Additional information is required to determine the evolution of this initial condition with respect to $z$ (i.e.\ the $z$ dependency of the function $B$), which is discussed below. 

\subsubsection{Second-order}\label{v2}

The continuity equation \eqref{eq:continuity} and the shear flow-rule equation \eqref{eq:flowshear}, at order $\delta ^ 2$, are
\begin{subequations}\label{eq:vel_second}\begin{align}  
\pdr{v^{(2)}}{y}+\frac{1}{h}\pdr{u^{(2)}}{n}&=-\pdr{u^{(1)}}{z},\\
\pdr{u^{(2)}}{y}+\frac{1}{h}\pdr{v^{(2)}}{n}&=-\pdr{v^{(1)}}{z} + 2\lambda^{(1)} \sigma^{(1)}_{xy}.
\end{align}\end{subequations}%
As for the stresses and the first-order velocities, these equations form a coupled wave equation for $u^{(2)}$ and $v^{(2)}$.
Following the same procedure as for the second order stresses, solving for $u^{(2)}$ and $v^{(2)}$ using the solutions for $u^{(1)}$, $v^{(1)}$, $\lambda^{(1)}$, and $\sigma_{xy} ^{(1)}$ from equation \eqref{eq:vel_first}, \eqref{eq:lambda_first}, and \eqref{eq:p_shear_first} gives a wave-like solution for $u^{(2)}$ and $v^{(2)}$.  As detailed in appendix~\ref{App:v_second}, this solution, when substituted into the boundary condition~\eqref{eq:boundary2 velocity} at $O(\delta^2)$, leads eventually (after significant but uninteresting algebra) to
\begin{multline}\label{eq:simplified_3}
\pdr{B(n+1,z)}{z}
- \frac{\intd h/\intd z}{h^2}A(n+1,z)
\\
- \left(\frac{1}{2}\frac{\intd p^{(0)}}{\intd z} - \frac{\intd h/\intd z}{h}\right)\!B(n+1,z)
+ \frac{1}{h}B'(n+1,z)A(n+1,z)
\\
 = -\frac{1}{h} \!\left( N (n+1,z) -  N (n-1,z)\right), 
\end{multline}
where $N(\xi,z)$ represents the complementary solution to the wave equation~\eqref{eq:vel_second}.  As for the stresses and the first-order velocities, setting the right hand side to zero to avoid $N$ growing as a function of $n$ gives the evolution of $B(\xi,z)$ in $z$.
This equation can be simplified by using $\alpha_1(z)$, $T(z)$ and $\omega(\xi,T)$ as before from~\eqref{eq:Burgers_total}, and by substituting
\begin{subequations} \begin{gather}
\alpha_2(z) = \exp\!\left\{\frac{1}{2}p^{(0)}\right\},
\qquad
Q(z) = \int _0 ^ z \frac{\alpha_1}{\alpha_2 h^2}\frac{\intd h}{\intd \tilde z} d \tilde z,
\\\Rightarrow\quad
\pdr{}{T} \!\left(\frac{B}{(\alpha_2 / h)}-Q\omega\!\right)\! + \omega  \pdr{}{\xi} \!\left(\frac{B}{(\alpha_2 / h)}-Q\omega\!\right)\!=0.
\label{eq:advection_w}  \end{gather} \end{subequations}
This is an advection equation with a velocity of $\omega$, which is the same advection velocity as the Burger's equation for the stresses~\eqref{eq:Burger}.  Equation \eqref{eq:advection_w} can be solved with a suitable initial condition derived from~\eqref{eq:v_ic}, and $B$ can then be calculated and used to find $u^{(1)}$, $v^{(1)}$ and $\lambda^{(1)}$ from~\eqref{eq:vel_first} and~\eqref{eq:lambda_first}.

\section{Computing the asymptotic solutions} \label{sec:computing}

Above we have derived governing equations for the various components of stress and velocity that can be integrated through the roll gap, either analytically or semi-analytically, starting from initial conditions at the roll-gap entrance and matching some conditions at the roll-gap exit.  
In this section, we summarise the solutions gained thus far and detail a simple numerical procedure for performing these calculations.  
We begin by considering the stresses, since they may be solved independently of the velocities.

\subsection{Stresses} \label{sec:sum_stress}
The stress profiles in terms of horizontal distance $x$ and vertical distance $y$ are written in terms of horizontal scales $n=\int_0^x 1/h\,\intd x$ and $z=\delta x$ as
\begin{subequations} \label{eq:sum_stress} \begin{gather} 
\label{eq:sum_p} p = p^{(0)}(z) + \delta \left[A\!\left(n+\frac{y}{h},z\right)+A\!\left(n-\frac{y}{h},z\right)\right] + O(\delta ^2)
\\
\label{eq:sum_xy} \sigma_{xy} = \delta \left[A\!\left(n+\frac{y}{h},z\right)-A\!\left(n-\frac{y}{h},z\right)+y\frac{\intd p^{(0)}}{\intd z}\right] + O(\delta ^2),
\\
\label{eq:sum_xx} \sigma_{xx} = 1-p,
\qquad\text{and}\qquad
 \sigma_{yy} = -1-p,
\end{gather}\end{subequations}%
where $A(\xi,z)$ is given below.  The leading-order pressure $p^{(0)}(z)$ is given by,
\begin{subequations} \label{eq:sum_p_lead}\begin{align} \label{eq:sum_p_lead_ent}
 \frac{\intd p^{(0)}}{\intd z} - \beta (1+p^{(0)}) -2\frac{\intd h}{\intd z}  &= 0, & &\text{for}\quad z<z_{\text{N}}
 \\\notag&&& \text{with} \quad p^{(0)}(z=0) = 1,
 \displaybreak[0]\\\label{eq:sum_p_lead_ex}
 \frac{\intd p^{(0)}}{\intd z}  + \beta (1+p^{(0)}) -2\frac{\intd h}{\intd z}&= 0, & &\text{for}\quad z>z_{\text{N}}
 \\\notag&&& \text{with} \quad p^{(0)}(z=1) = 1, 
 \end{align}\end{subequations}%
with $z_{\text{N}}$ chosen such that $p^{(0)}$ is continuous at  $z_{\text{N}}$. The solution for $p^{(0)}$ is chosen to satisfy the forward and backward tension conditions, which are taken to be zero for the results presented below.
Therefore,  $p^{(0)}$ is solved by integrating equation~\eqref{eq:sum_p_lead_ent} forward from the entrance, and integrating equation~\eqref{eq:sum_p_lead_ex} backwards from the exit, using the \Matlab\ ODE solver \texttt{ode45} \citep{MATLAB:R2024a_u1}. 
This is the same solution as the slab method and the two curves thus produced are referred to as the pressure hill.

The first-order wave-like oscillatory function $A(\xi,z)$, in the set of equations~\eqref{eq:sum_stress}, obeys Burger's equation once suitably rescaled,
\begin{subequations} \label{eq:sum_Burger}\begin{align} 
A(\xi,z) &= \frac{\alpha_1(z)}{h(z)} \omega\big(\xi,T(z)\big),
\quad
T(z) = \int^z \!\!\!\frac{\alpha_1}{h(\bar z)^2}\ \mathrm{d} \bar z,
\\
\alpha_1(z) &= \exp \Bigg\{\! \int ^ z _0 \Big( \mp \frac{\beta}{2h(\Tilde{z})}(p^{(0)}(\Tilde{z}) - 1) \Big) \intd\Tilde{z} \Bigg\},
\\
\pdr{\omega}{T} &+ \frac{1}{2}  \pdr{}{\xi} \Big( \omega^ 2 \Big) = 0 \qquad\text{for}\quad -1<\xi<1,
 \end{align}\end{subequations}%
with $A$ and therefore $\omega$ periodic in $\xi$ such that $A(\xi+1,z) = A(\xi-1,z)$; hence, it is sufficient to solve Burger's equation for $-1 < \xi < 1$. Burger's equation \eqref{eq:sum_Burger} is solved first from the entrance at $z=0$ to the neutral point at $z=z_{\text{N}}$ to give $A^{(-)}$, and then again from $z=z_{\text{N}}$ onwards with the new initial condition at the neutral point to give $A^{(+)}$:
\begin{subequations}\label{eq:sum_A}\begin{align}
A(\xi,z) &= 
\begin{cases} 
A^{(-)}(\xi,z) &\displaystyle \text{for }z < z_{\text{N}}
\\
A^{(+)}(\xi,z) &\displaystyle \text{for }z > z_{\text{N}} 
\end{cases}
\\
A^{(-)} \left(\xi,\,0\right) &= \frac{1}{2}\!\left( \tau_{0}(\xi) - \sigma_{xx}^{(1)}(\xi,0) \right) -\frac{1}{2} \xi\!\left.\frac{\intd p^{(0)}}{\intd z}\right|_{\mathrlap{z=0}}
\\ 
A^{(+)}(\xi,z_{\text{N}}) &= A^{(-)}(\xi,z_{\text{N}}) + \beta\xi\big(p^{(0)}(z_{\text{N}})+1\big).
\end{align}\end{subequations}%
For the initial condition at $n=z=0$ in \eqref{eq:sum_A}, the simplest assumption is that $\sigma^{(1)}_{xx}$ is zero and  $\tau _ 0 (y)$ is linear in $y$ at the entrance; taking $\tau_0(y)=-1.2y$ was found to give a good comparison for all rolling parameters considered here. 
This implies all material points are affected the same by the inlet tension, and the shear stress linearly increases from zero at the symmetry line to a maximum at the surface.
While this maximum shear stress can be chosen to give good agreement with FE results, as will be seen below, an improved estimation is obtained by allowing $\sigma^{(1)}_{xx}$ to vary quadratically (but with zero average to maintain the same inlet tension according to \eqref{eq:boundary tension}). 
Both scenarios are compared with FE results in Section~\ref{sec:result}. 
Once $A^{(-)}$ is completely solved from \eqref{eq:sum_Burger} with the initial conditions at $n=z=0$, the solution at $z_{\text{N}}$ is used to modify the initial condition at the neutral point. 
This new initial condition is then used to solve $A^{(+)}$ from the neutral point to the exit. 
$A^{(-)}$ and $A^{(+)}$ together give the full solution in the entire roll gap. 

Burger's equation  \eqref{eq:sum_Burger} is solved with the finite volume method in \Matlab, and a limiter is applied to ensure that the numerical solution does not develop new extrema. 
While the periodicity of $A$ in $\xi$ means that $A$ will have discontinuities (e.g.\ at $\xi=\pm 1$), these discontinuities form an expansion fan in the Burger's equation solution, and so are not problematic. Using the unoptimised \Matlab\ code provided in the supplementary material, solving for all stresses takes a couple of seconds on a standard laptop.

\subsection{Velocity} \label{sec:sum_vel}
The velocity profiles are given by,
\begin{subequations} \label{eq:sum_vel}\begin{align} 
u &= \frac{1}{h} + \delta\! \left[ B\!\left(n+\frac{y}{h},z\right) +B\!\left(n-\frac{y}{h},z\right) \right]\! + O(\delta ^2),\\
v &= - \delta\! \left[B\!\left(\!n+\frac{y}{h},z\!\right) -B\!\left(\!n-\frac{y}{h},z\!\right)\! - y  \frac{\intd h/ \intd z}{h^2}\right]\! + O(\delta ^2),
\end{align}\end{subequations}%
where the leading-order slab-like behaviour is controlled directly by the roll-gap thickness imposed by the rolls, and the wave-like oscillatory correction obeys,
\begin{subequations} \label{eq:sum_advection}\begin{gather} 
\pdr{}{T} \left(\frac{B}{(\alpha_2 / h)}-Q\omega\right) + \omega  \pdr{}{\xi} \left(\frac{B}{(\alpha_2 / h)}-Q\omega\right)=0,
\\
Q(z) = \int _0 ^ z \frac{\alpha_1}{\alpha_2 h^2}\frac{\intd h}{\intd \tilde z} d \tilde z,
\quad
\alpha_2(z) = \exp\left\{\frac{1}{2}p^{(0)}\right\},
\\
\text{with}\qquad B(\xi,\,0) =\frac{\xi}{2}\!\left.\frac{\intd h}{\intd z}\right|_{\mathrlap{z=0}},
 \end{gather}\end{subequations}%
for $-1<\xi<1$ and with periodicity $B(\xi+1,z) = B(\xi-1,z)$ otherwise, as for $A(\xi,z)$.

Similarly to $A(\xi,z)$, the advection equation \eqref{eq:sum_advection} may be fully solved for $B(\xi,z)$ using suitable initial conditions at $z=0$ for $-1 < \xi < 1$, with discontinuities at the beginning and end of each interval forced by the periodic nature of $B(\xi,z)$, similar to $A(\xi,z)$, which again results in an expansion fan.
However, a subtlety of the solution here, unlike the Burger's equation solution for $A(\xi,z)$, is that a discontinuous initial condition coupled with the advection equation for $B(\xi,z)$ does not completely determine the solution within the expansion fan. 
This distinction arises because, in Burger's equation, the function within an expansion fan is inversely proportional to the slope of characteristics, but this does not apply to the advection equation.  
Consequently, additional information is needed to specify the initial condition for the expansion fan for $B(\xi,z)$. 
Physically, this might come from solving an inner elastic problem at the contact point at the entrance, a topic not addressed in this study. 
Consequently, here we assume that, as for Burger's equation, the expansion fan is initially linear (consistent with our assumption of rigid body motion at the entrance), and instead of specifying $B(\xi,z)$ at $z=0$ and integrating forwards, we specify $B(\xi,z)$ for a small nonzero value of $z$ close to the entrance and solve for the evolution forward to the exit and backward to the entrance using the advection equation~\eqref{eq:sum_advection}. 
This allows us to use a continuous initial condition for our solver, chosen such that the solution at the entrance is the desired one.

The numerical solution to \eqref{eq:sum_advection} is non-trivial, with the advected velocity depending on both $\xi$ and $z$ and being positive and negative. 
Here, an upwinding explicit finite difference method is employed, with the step size varied in order to maintain a CFL constraint. The results plotted here are obtained in a few seconds using unoptimised \Matlab\ code on a standard laptop.

\section{Results and comparison with FE simulation}\label{sec:result}

Since it is extremely difficult to experimentally observe the stress pattern through thickness during the rolling process, the predictions of the asymptotic model are here compared with carefully conducted FE simulations using the \Abaqus\ package; full details are given by~\citet{flanagan2024}. 
In summary, simulations are made in \Abaqus/Standard with an implicit solver. 
Only one roll and half the sheet are simulated following the assumption of symmetry about the sheet’s horizontal centre plane. 
To comply with the plane-strain assumption, the sheet metal is modelled as a two-dimensional deformable part and the rolls as a two-dimensional analytical rigid geometry. 
The contact between the roll and sheet is discretised using the surface-to-surface method. 
Since a mesh sensitivity study confirmed that under-resolution through-thickness strongly affects the results~\citep{flanagan2024},  
here, 30 CPE4R elements were used through the half-thickness of the sheet; this showed the best trade-off between accuracy and computation time~\citep{flanagan2024}. 
The simulation consists of two steps: the bite step, where the roll is slowly translated vertically to indent the sheet; and the rolling step, where the sheet is horizontally displaced due to rotation of the roll. 
By carefully initiating the simulation with this separate bite step, initial transients and numerical oscillations commonly observed in this type of FE calculation are minimised without the need to resort to a custom-written bespoke numerical solver (for example employing a mixed Eulerian--Lagrangian formulation).
The model is then run for a sufficiently long time to attain a steady state, as measured by obtaining steady through-thickness stress and strain distributions, rather than simply by observing roll force and torque \citep{flanagan2024}. 
For static \Abaqus/Standard simulations, although stresses are generated directly, velocity must be calculated from the displacement change between time frames. 
The computation time for these simulations depends on the aspect ratio, and an example of $\delta = 0.125$ requires a computation time of 15.89 hours.

For the simulations, the sheet material is chosen to be a mild steel (grade DC04), a common material used in cold metal forming~\citep{spittel2009steel}. 
While the mathematical model uses $\hat \kappa = 288.67\,\mathrm{MPa}$ (corresponding to $\hat Y = 500\,\mathrm{MPa}$), the FE simulations were performed with two different material models: 
first with the yield stress fixed at $\hat \kappa = 288.67\,\mathrm{MPa}$, so that the simulation is close to the perfectly-plastic material assumed in the mathematical model; and then with a more typical hardening profile where the initial yield stress of $\hat \kappa = 275.51\,\mathrm{MPa}$ rises to $324.29\,\mathrm{MPa}$ at a true plastic strain of $0.4$ as a result of strain hardening. 
The results presented below are for the non-hardening material model unless otherwise stated.  
A realistic Young’s modulus of $E=206.3\,\mathrm{GPa}$ and a Poisson ratio $\nu = 0.3$ are used for all FE simulations.

For the FE simulations, the rolls rotate with a surface speed of $1.28\,\mathrm{ms^{-1}}$, the sheet has an initial full thickness $2\hat h_0 = 4\,\mathrm{mm}$, and the size and position of the rolls is varied to give the required roll-gap aspect ratio and a reduction of $r=0.25$. 
The friction coefficient between the workpiece and the rolls is chosen to be $0.1$, except for short roll gaps with $\delta = 0.3$ and $0.5$, for which $\mu = \delta/2$ is chosen to guarantee the initial bite.  
These choices of material, friction coefficient, initial thickness, reduction, and roll speed are purely illustrative, and solutions may be found for any values of these parameters.

\subsection{Prediction of through-thickness variation} \label{sec:field variation}
\begin{figure*}[t] %
\centering %
\includegraphics{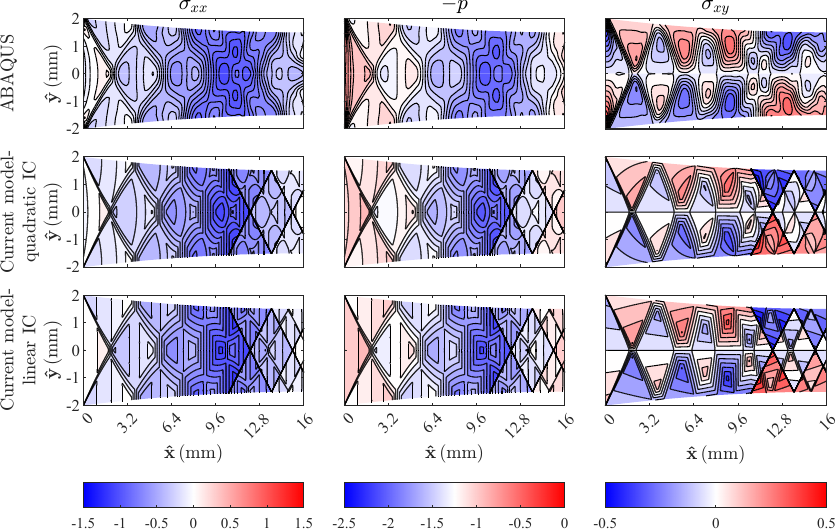}%
\caption{Comparison of results for $\delta=0.125$ from FE simulations~\citep{flanagan2024} (top row) with the current model (equation~\ref{eq:sum_stress}), with two different initial conditions: quadratic initial condition (middle row) and linear initial condition (bottom row). Left: contour plots of horizontal stress. Middle: contour plots of pressure. Right: contour plots of shear stress. Results are shown in dimensionless form, i.e. scaled with $\hat{\kappa}$, the yield stress in shear. Parameters used are $(\hat h_0, r, \mu,\hat \kappa)=(2\,\mathrm{mm}, 25\%, 0.1, 288.67\,\mathrm{MPa} )$.}%
\label{fig:effect of ic}%
\end{figure*}%
Figure~\ref{fig:effect of ic} compares the axial stress $\sigma_{xx}$, hydrostatic pressure $-p$, and shear stress $\sigma_{xy}$ predicted by the \Abaqus\ simulation and the mathematical model for $\delta=0.125$ with zero backwards/forwards tensions. 
The top row shows FE simulation and the two bottom rows show results from the mathematical model with two different inlet boundary conditions. 
According to the \Abaqus\ simulation, even with zero back tension, the distribution of $\sigma_{xx}$ is non-zero at the entrance, which can be attributed to elastic deformation. 
If known, this can be included in the mathematical model by using a quadratic function as an initial condition as explained in \ref{sec:sum_stress}, and the result of this is shown in the middle row in Figure~\ref{fig:effect of ic}. 
In this figure, the unknown term in~\eqref{eq:sum_A} is taken to be $\tau_{0}(y) - \sigma_{xx}^{(1)}(y,0) = -2(0.15y^2+0.5y-0.05)$, where the coefficients are chosen by comparing the results with simulation and imposing a zero average \eqref{eq:boundary tension}. 
However, reasonable results can still be obtained by disregarding the initial distribution of $\sigma_{xx}$ and imposing a linear distribution for shear stress  $\tau _0$ as the initial condition (see the bottom row in Figure~\ref{fig:effect of ic}; for this case, $\tau _0=-1.2y$ was chosen, which gives a good comparison for all rolling parameters considered here).
This value can be utilised if the model is intended to be employed in a fully predictive way without relying on FE data.
Comparing the two bottom rows in Figure~\ref{fig:effect of ic} also demonstrates that, with the cost of an extra degree of freedom, the quadratic initial condition results in a rounded pattern which is more similar to the FE simulation. 
The mathematical model's prediction of stresses at the roll-gap exit differs from those of the FE simulation due to the workpiece unloading and becoming purely elastic, although this appears to be localised to the roll-gap exit and does not appear to affect the agreement of the solution elsewhere within the roll gap.

The oscillatory pattern in the pressure distribution in Figure~\ref{fig:effect of ic} is very similar to $\sigma_{xx}$, and with a similar effect of the initial conditions at the roll-gap entrance, as is expected from equations~\eqref{eq:sum_xx}. 
The pressure is the sum of the leading-order solution describing the ``pressure hill'' and the $O(\delta)$ correction term causing the oscillatory pattern. 
The pressure hill causes the pressure to approximately increase in magnitude up to the neutral point ($\hat x \approx 10.7$) and decrease after that. 
Nevertheless, this trend is not monotonic within either of the two zones due to the correction term; each momentary increase in pressure is followed by a subsequent decrease, resulting in the formation of multiple local peaks.  
This means that, unlike the slab method prediction, the maximum pressure does not necessarily occur exactly at the neutral point.

\begin{figure*}%
\centering%
\includegraphics[width=\linewidth,height=0.9\textheight,keepaspectratio]{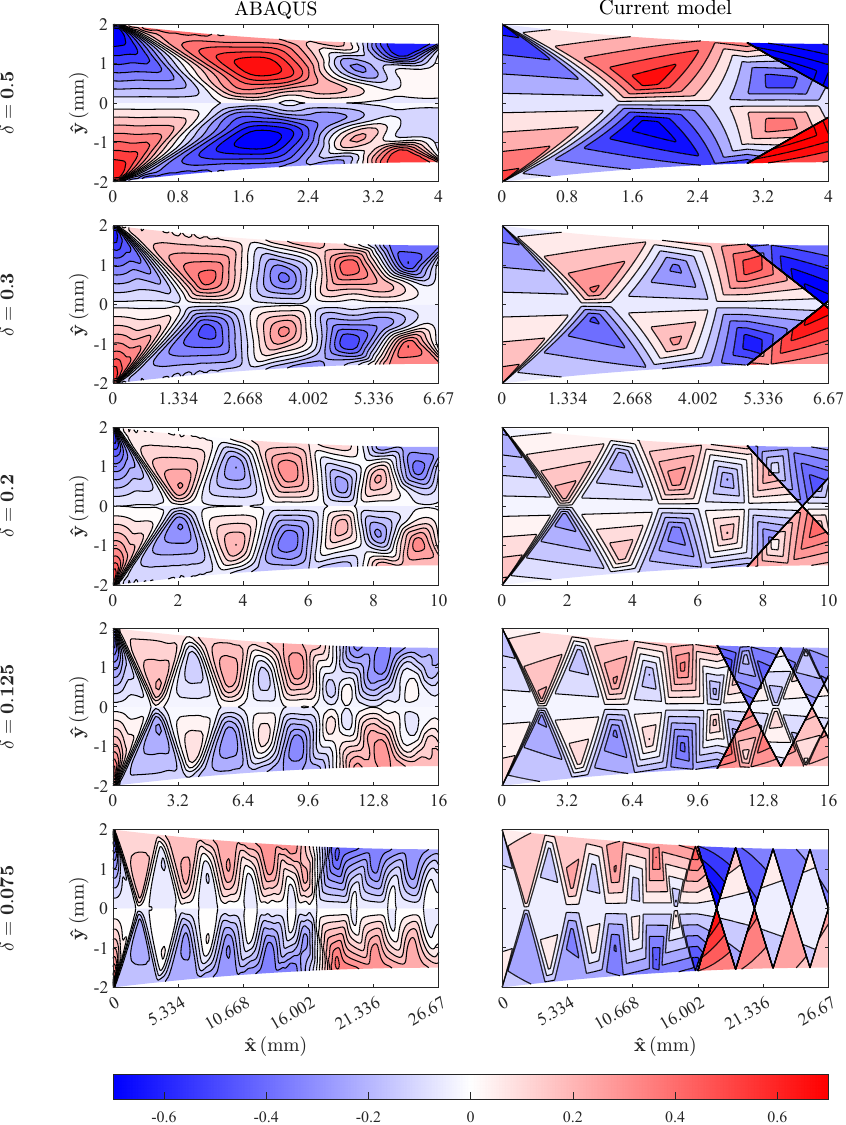}%
\caption{Comparison of shear stress for varying $\delta$ from the current model (right, equation~\eqref{eq:sum_xy}) and FE simulations~\citep{flanagan2024} (left). Stresses are scaled with $\hat{\kappa}$, the yield stress in shear. Parameters used are $(\hat h_0,r,\mu,\hat \kappa)=(2\,\mathrm{mm}, 25\%, 0.1, 288.67\,\mathrm{MPa})$, except $\delta=0.5$ uses $\mu=0.25$ and $\delta=0.3$ uses $\mu=0.15$.}%
\label{fig:shearcom}%
\end{figure*}

The shear stress in the last column in Figure~\ref{fig:effect of ic} is also well replicated by the mathematical model. 
The different initial conditions in the last two rows in Figure~\ref{fig:effect of ic} do not significantly affect the shear distribution pattern.
This implies that a linear variation of shear stress is sufficient to model initial condition in~\eqref{eq:sum_A} for the remainder of this paper.
Unlike many other mathematical models, where the application of Coulomb friction is associated with a vertical discontinuity at the neutral point (as illustrated in Figure~\ref{fig:slab}), the current model correctly predicts the trend of the changing sign in shear along the diagonal lines. 
This is shown more clearly for shorter roll gaps, plotted as the top rows in Figure~\ref{fig:shearcom},
where the shear stress fields from FE (left column) and the mathematical model (right column) are plotted for varying $\delta=\hat h_0/\hat l$.

The discontinuity in shear stress at the surface, observed in the mathematical model in Figures~\ref{fig:effect of ic} and~\ref{fig:shearcom}, is an unavoidable consequence of our assumption of Coulomb friction without sticking. The shear stress discontinuity at the surface for the mathematical model causes a shock to form in the Burger's equation solution, producing the sharp line starting from the neutral point. The FE simulation, on the other hand, predicts a sticking region, giving a smoother transition between positive and negative shear at the neutral point than is observed in the mathematical model. Still, the predicted neutral point is correctly positioned compared with the FE simulations for all aspect ratios.

Another feature seen in Figures~\ref{fig:effect of ic} and~\ref{fig:shearcom} is the initial diagonal discontinuity in stresses emanating from the first contact point between the workpiece surface and the rolls (e.g.\ the top left and bottom left points in each plot); this initial discontinuity subsequently spreads out and weakens.
This is the result of the evolution of the discontinuity in the initial condition (as explained in~\ref{sec:sum_stress}) as an expansion fan. 
This field is quite common in plane-strain problems at die corners or at the location of sudden changes in cross-section solved by the slip-line method \citep{johnson1982plane}, and is also well predicted by the current method.  
Significant numerical resolution at the contact points is needed for FE simulations to accurately reproduce the same behaviour.

\begin{figure*}%
\centering %
\includegraphics{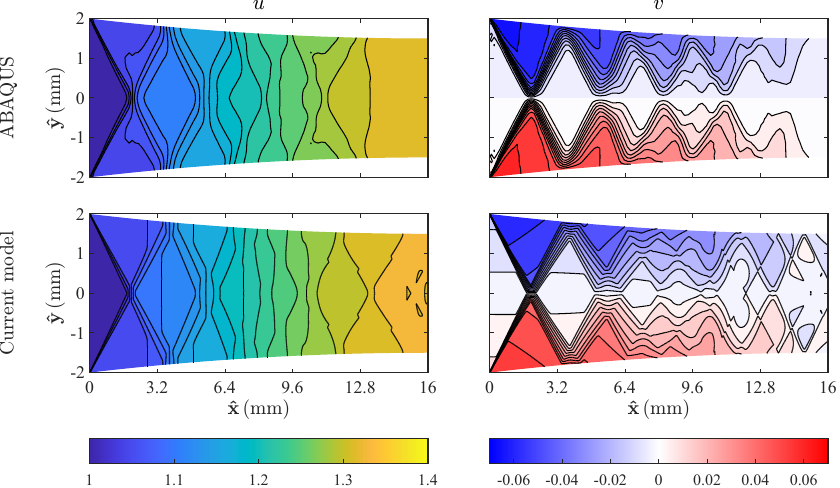}%
\caption{Comparison of horizontal (left) and vertical (right) velocities from the current model (equation~\eqref{eq:sum_vel}) (top row) with FE simulations~\citep{flanagan2024} (bottom row) for $\delta=0.125$. Velocities are scaled with $\hat U_0$, the entrance velocity. Other parameters used are  $(\hat h_0, r, \mu,\hat \kappa)=(2\,\mathrm{mm}, 25\%, 0.1, 288.67\,\mathrm{MPa} )$}%
\label{fig:velocities}%
\end{figure*}

The trend observed in Figure~\ref{fig:shearcom} indicates that the oscillatory pattern is not restricted to a particular aspect ratio, as similar trends are seen for all values of $\delta$. 
The number of lobes present increase with roll-gap length and are approximately proportional to the aspect ratio of the roll gap $1/\delta$. 
This may be explained by the deformation mechanism in the roll gap being effectively that of uniaxial extension, with the sheet getting longer and thinner, although forced by the rolls rather than extensive normal forces at the entrance and exit.  
This means the slip lines, along which information about the deformation is carried, align with the local direction of maximum shear at 45 degrees to the centre line.  
As information only reaches a material element along slip lines, information about initial contact with the rolls, friction, the neutral point, etc, all travel at 45 degrees to the centre line.
This sets up an oscillatory pattern that repeats, because of the 45 degree angle, on average every sheet thickness, meaning the number of oscillations depends on the number of sheet thicknesses that fit into the length of the roll gap.
This becomes more comprehensible when viewed alongside the velocity distribution in Figure~\ref{fig:velocities}.

Figure~\ref{fig:velocities} shows the vertical and horizontal velocity distribution for $\delta=0.125$ both from FE simulation and the mathematical model. The results from the current model with the assumption of a homogeneous rigid-body velocity profile at the entrance align well with FE simulations, despite its elasto-plastic material model. The agreement is attributed to simulating a single pass where the sheet enters horizontally into the roll gap.
The frequency of the pattern in vertical and horizontal velocity is almost the same as for the stresses (Figure~\ref{fig:shearcom}). 
In the vertical velocity contour plots in Figure~\ref{fig:velocities}, as expected, the material points are pushed towards the centre line due to the presence of the rolls, although interestingly this is limited to certain regions, between which there is almost no vertical velocity.  
Each zone exhibits distinct horizontal and vertical velocities that are suggestive of a block gliding over its neighbouring block; this provides another alternative interpretation of the shear pattern observed in Figure~\ref{fig:shearcom}.

\begin{figure*}[t]%
\centering %
\includegraphics{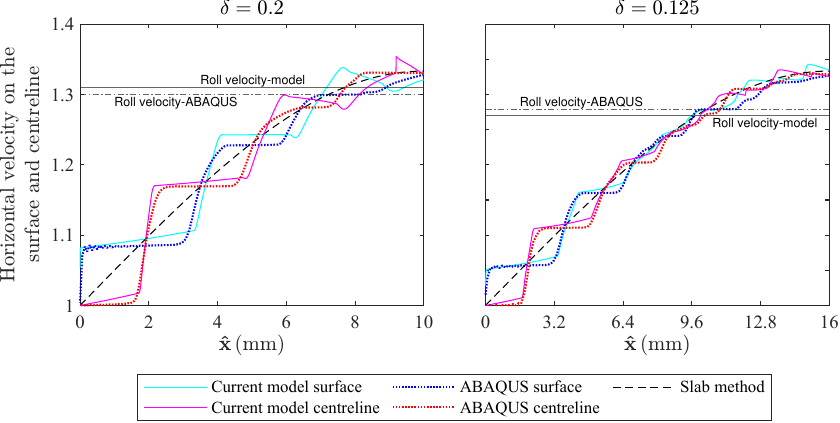}%
\caption{Plots of horizontal sheet velocity $u$ on the roll surface and along the centreline against distance through the roll gap $\hat{x}$, for $\delta=0.2$ (left) and $\delta=0.125$ (right). Velocities are scaled with $U_0$, the entrance velocity. Other parameters used are  $(\hat h_0,r,\mu,\hat \kappa)=(2\,\mathrm{mm}, 25\%,0.1,288.67\,\mathrm{MPa})$.  Also plotted are the horizontal roll-surface velocities for comparison.}%
\label{fig:velsurface}%
\end{figure*}

For the horizontal velocity, the leading order $u^{(0)}$, and indeed the classic slab method, both predict a smoothly increasing horizontal velocity from the entrance to the exit. 
However, including the correction term gives the horizontal velocity shown in Figure~\ref{fig:velocities} characterised by a series of incremental steps.  This is more evident in Figure~\ref{fig:velsurface}, where the horizontal velocity along the surface and along the centreline are compared for the slab method, the current model, and FE simulations.
The sheet is introduced into the roll gap horizontally at a velocity lower than that of the rolls, and initially, while the surface accelerates after contact with the roll, the centre maintains its initial velocity. 
This trend then reverses and the centre line advances more rapidly than the surface. 
The leapfrogging of surface and centerline continues up to some point near the exit, where the entire block travels uniformly and exits the roll gap in a horizontal direction. 
Also plotted in Figure~\ref{fig:velsurface} are the horizontal roll-surface velocities for comparison.  In particular, the \Abaqus\ roll-surface and sheet-surface velocities coincide over a sticking region, while the model sheet-surface velocity increases past the the roll-surface velcocity at the neutral point, as expected from the differences in sticking and slipping behaviour between the model and the FE discussed when considering the stresses above.

\begin{figure*}[t]
    \centering
    \includegraphics{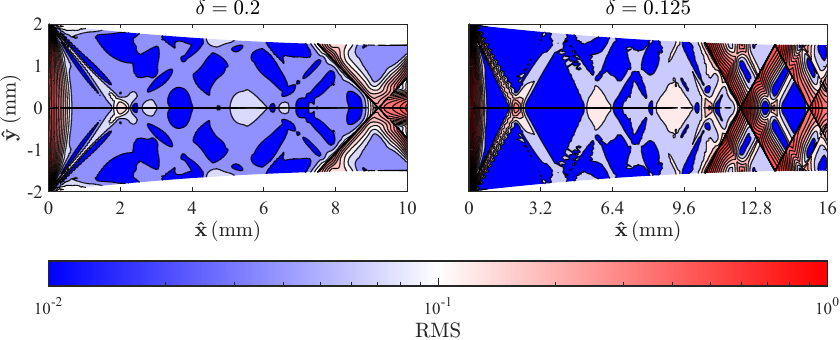}
    \caption{RMS differences between the FE and model stresses (combining $\sigma_{xx}$, $\sigma_{yy}$ and $\sigma_{xy}$) for $\delta=0.2$ (left) and $\delta=0.125$ (right).}
    \label{fig:rms_contour}
\end{figure*}

\begin{figure}
    \centering
    \includegraphics[width=\linewidth,height=0.25\textheight,keepaspectratio]{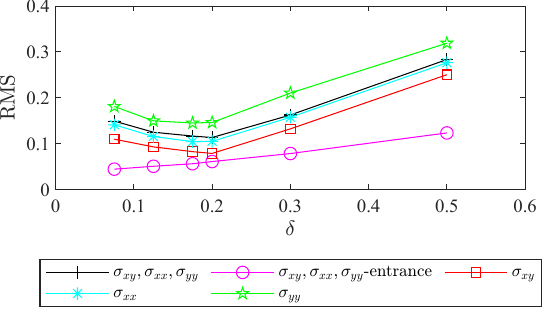}
    \caption{Averaged RMS differences between the FE and model stresses for different values of $\delta$ and different stress components ($\sigma_{xx}$, $\sigma_{yy}$, $\sigma_{xy}$, and all combined).  The ``entrance'' values are calculated only from the entrance to the neutral point.}
    \label{fig:rms_delta}
\end{figure}

By way of a more quantitative comparison between the FE and model stresses, figure~\ref{fig:rms_contour}
plots the root--mean--square (RMS) differences of the $\sigma_{xx}$, $\sigma_{xy}$ and $\sigma_{yy}$ stress components between the FE results and the model results.  Before the neutral point, the difference is mostly $O(\delta^2)$, as expected from the asymptotic analysis; however, after the neutral point, the difference increases to $O(\delta)$ due to the assumption of no sticking region in the model.  Significant differences are also seen at the entrance owing to the initial deformation in the FE results being below-yield and elastic, while the model neglects elasticity.  The average RMS difference is dominated by the larger post-neutral-point differences, as is demonstrated in figure~\ref{fig:rms_delta}.
The average RMS differences for each component of the stress follow the same trend as the combined differences, and are significantly larger than the average differences in the entrance region only (averaged from the entrance up to the neutral point).  The asymptotics is expected to give more accuracy for smaller values of $\delta$, and this is seen in the entrance region.  However, the differences between the FE and the model increase beyond the neutral point as $\delta$ is reduced, owing to the increased number of oscillations in this region for smaller $\delta$.

\subsection{Comparison to slab analysis}

\begin{figure*}%
\centering%
\includegraphics{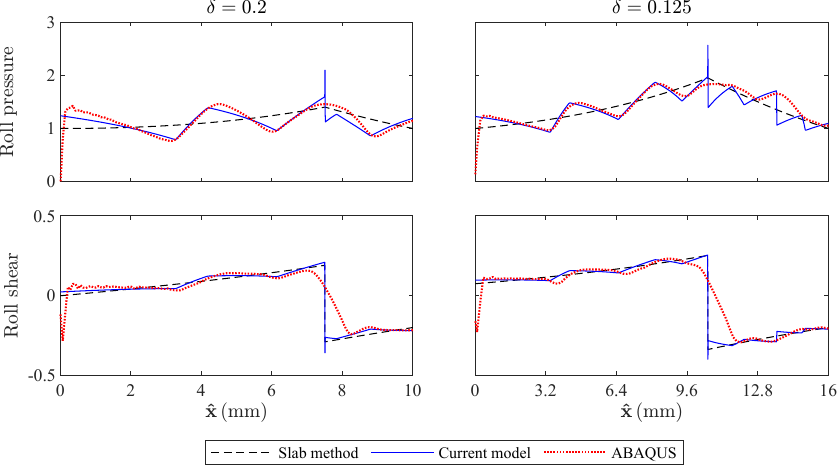}%
\caption{Plots of $p$ (top) and $\sigma_{xy}$ (bottom) on the roll surface, against distance through the roll gap $\hat{x}$, for $\delta=0.2$ (left) and $\delta=0.125$ (right). Stresses are scaled with $\hat{\kappa}$, the yield stress in shear.  Other parameters used are  $(\hat h_0,r,\mu,\hat \kappa)=(2\,\mathrm{mm}, 25\%,0.1,288.67\,\mathrm{MPa})$.}%
\label{fig:rollpressure}%
\end{figure*}

Since slab analysis is not intended to give through-thickness information, it would be unfair to compare the results above with the slab method.
Instead, here we use the stress fields on the roll surface to compare the distribution of roll pressure and roll shear between FE, slab method, and the current model. Results for two aspect ratios are plotted in Figure~\ref{fig:rollpressure}. 
The \quotes{slab method} roll pressure is calculated using the leading-order pressure~\eqref{eq:sum_p_lead}, and the \quotes{current model} roll pressure is~\eqref{eq:sum_p}. 
For roll shear, the \quotes{slab method} is $y\,\intd p^{(0)}\!/\intd z$, 
 and the $O(\delta ^2)$ correction given by $\sigma_{xy}^{(2)} \mp \beta p^{(1)} = 0$ at $y=h$ is added to get the \quotes{current model} roll shear.

The mathematical model accurately captures the oscillations in roll pressure observed in the FE data by incorporating a correction term of magnitude $\delta$.  
For short roll-gap lengths, this correction is not necessarily small, and in fact, the successive pressure peaks completely overcome the classic slab method pressure hill profile, as was observed experimentally by \citet{al1973experimental}. 
These deviations are unrelated to elastic deformation, roll flattening, or hardening, as those effects are not included here. 
As the sheet length increases, the stress oscillations become small fluctuations to the pressure hill.  
These oscillations average out when integrating the roll pressure and shear over the whole roll surface to produce the roll force and roll torque; this is why roll force and torque are poor measures of the accuracy of a rolling simulation~\citep{flanagan2024}, and why classical slab methods produce reasonable predictions for the total roll force and torque. 

When considering shear, the current model prediction is identical to the slab method to the first-order correction on the surface (owing to equation~\ref{eq:sum_xy} and periodicity in $A$). 
However, incorporating the $O(\delta ^2)$ correction on the surface accounts for the oscillations detected in the simulation, and this is what is plotted as \quotes{Current model} roll shear in Figure~\ref{fig:rollpressure}.
The number of local peaks increases for longer sheet lengths, as expected from Figure~\ref{fig:shearcom}. 
As mentioned in the previous section, the observed discontinuity on the surface at the neutral point is an inevitable consequence of slipping Coulomb friction, while the FE simulation varies smoothly due to a small zone of sticking friction. 
This could be improved in future by incorporating sticking friction in the asymptotic model, although a more rigorous study would be required also incorporating elasticity, as the friction rule alone is not solely responsible for the behaviour around the neutral point~\citep{orowan1943calculation,flanagan2024}.  Such alternative friction treatments are beyond the scope of the present study. 
Also, the sudden changes observed near the entrance in the FE roll pressure and roll shear results are attributable to elastic deformation, which is not accounted for in the current model.

\subsection{Comparison with a work-hardening FE simulation}
\label{sec:work-hardening}

\begin{figure*}[t]%
\centering %
\includegraphics{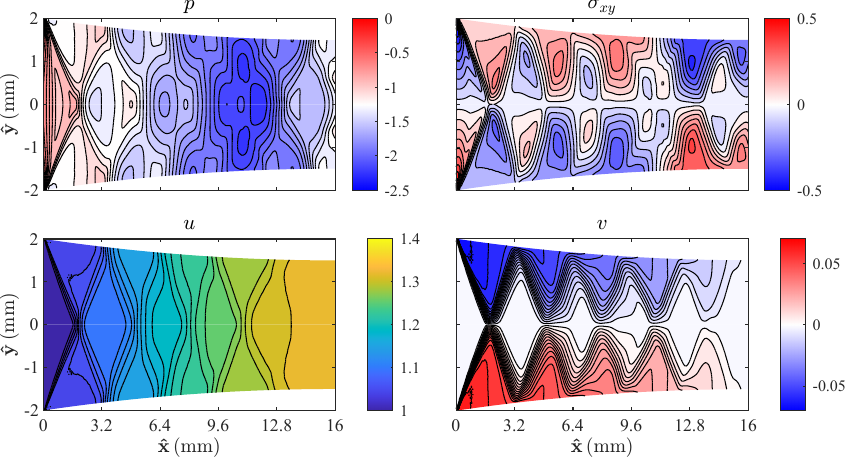}%
\caption{Contour plots of FE results with strain hardening~\citep{flanagan2024} for $\delta=0.125$. The initial yield stress $\hat \kappa=275.51\,\mathrm{Mpa}$ increasing to $324.29\,\mathrm{MPa}$ at a true plastic strain of $0.4$. Results are scaled with the initial yield stress $\hat \kappa=275.51\,\mathrm{Mpa}$. Other parameters used are $(\hat h_0, r, \mu)=(2\,\mathrm{mm}, 25\%, 0.1)$.}%
\label{fig:hardening}%
\end{figure*}

Up to now, we have compared against an FE simulation using a non-hardening material with a fixed yield stress of $\hat \kappa = 288.67\,\mathrm{MPa}$.  
Here, FE results are presented for a more realistic hardening material: 
the initial yield stress is taken to be $\hat \kappa = 275.51\,\mathrm{MPa}$, increasing to $324.29\,\mathrm{MPa}$ at a true plastic strain of $0.4$ as a result of strain hardening.  
Strain-hardening simulations were performed for the whole range of aspect ratios shown in Figure~\ref{fig:shearcom}, and although we only present results here for $\delta=0.125$ for brevity, these results are typical for all values of $\delta$.
Contour plots of stress and velocity for the hardening material for $\delta=0.125$ are shown in Figure~\ref{fig:hardening}.
Comparing to the equivalent non-hardening results in Figures~\ref{fig:effect of ic} and~\ref{fig:velocities}, hardening is seen to cause a progressive increase in the magnitude of stress from left to right as the workpiece passes through the roll gap, but to have a negligible effect on the through-thickness oscillatory pattern.

We next compare the current non-hardening mathematical model against FE simulations of this hardening material.
\begin{figure*}[t]\centering
\includegraphics{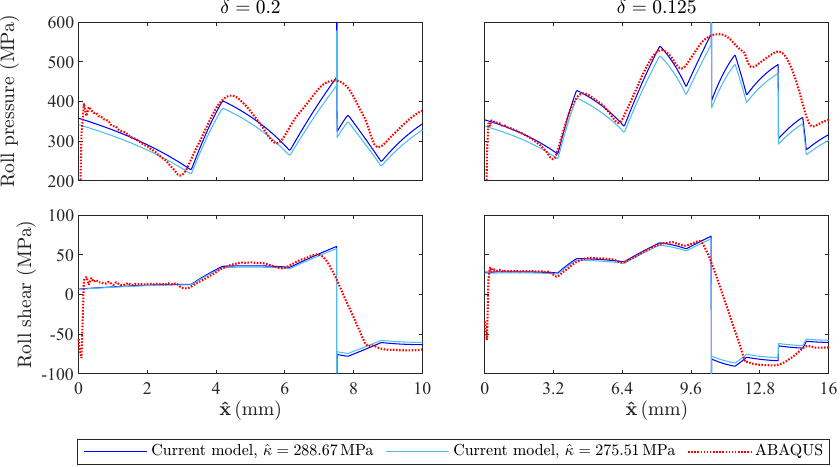}%
\caption{Plots of $p$ (top) and $\sigma_{xy}$ (bottom) on the roll surface, against distance through the roll gap, for (left) $\delta=0.2$ and (right) $\delta=0.125$. \Abaqus\ results~\citep{flanagan2024} include hardening, with an initial yield stress of $\hat \kappa = 275.51\,\mathrm{MPa}$ increasing to $324.29\,\mathrm{MPa}$ at a true plastic strain of $0.4$. The current non-hardening model is plotted for two different yield stresses.  Other parameters used are  $(\hat h_0,r,\mu)=(2\,\mathrm{mm}, 25\%,0.1)$ } %
\label{fig:roll_hardening}%
\end{figure*}%
Figure~\ref{fig:roll_hardening} shows pressure and shear distribution on the sheet surface from the current model and from FE simulation. 
Since in the mathematical model all stresses are scaled by the yield stress $\hat \kappa$, the prediction depends on its magnitude. 
To see the effect, we compare the simulation with the mathematical model twice: once using the initial yield stress $\hat \kappa = 275.51\,\mathrm{MPa}$; and once using the average yield stress from the FE simulation, $\hat \kappa = 288.67\,\mathrm{MPa}$. 
The impact of the different stress scalings is more noticeable on the pressure rather than shear stress, and overall, the larger scaling ($\hat \kappa = 288.67\,\mathrm{MPa}$) results in better agreement. 
The agreement is understandably better at the start of the roll gap, when the sheet undergoes little hardening, and becomes progressively worse as more hardening takes place. 
In summary, this comparison between the hardening simulation and non-hardening model demonstrates that the stress oscillations are unaffected by the hardening characteristics, and that the existing model remains a useful predictor for hardening materials. 
Further improvement might be achieved in future by explicitly incorporating hardening into the mathematical model.

\section{Appraisal of simplifying assumptions}
\label{sec:assumptions}

The systematic assumptions used in the model along with their justifications are summarized in table~\ref{table:assumptions}.
\begin{table*}%
\caption{Assumptions used in this work along with their justifications. }
\centering
\vspace{5pt}
\begin{tabular}{ |>{\centering\arraybackslash}p{5.5cm}||>{\centering\arraybackslash}p{3cm}|>
{\centering\arraybackslash}p{1.1cm}|>
{\centering\arraybackslash}p{1.1cm}|>
{\centering\arraybackslash}p{1.1cm}|}
 \hline
\textbf{Assumption} & \textbf{Justification} & \textbf{Slab} &  \textbf{Maths} &  \textbf{FE} \\
\hline
\hline
Plane strain & common\newline(wide sheet)  & \cmark & \cmark & \cmark \\ \hline
Rigid rolls & common\newline(not foil) & \cmark & \cmark & \cmark \\ \hline
Isotropy (von Mises) & simplicity & \cmark & \cmark & \cmark \\ \hline
Steady state & agrees with FE & \cmark & \cmark & \xmark \\ \hline
No elasticity& common & \cmark & \cmark & \xmark \\ \hline
No plastic hardening & simplest model  & \cmark & \cmark & \xmark \\ \hline
Large $\hat \ell/ \hat h$ & thin sheet rolling\newline agrees with FE  & \xmark & \cmark & \xmark \\ \hline
Small friction & lubricated rolling  & \xmark & \cmark & \xmark \\ \hline
Slipping everywhere &  simplest model & \xmark & \cmark & \xmark \\ \hline
Homogenous velocity ($u^{(1)}$ and $v^{(1)}$ are zero) at the entrance&  agrees with FE  & \xmark & \cmark & \xmark \\ \hline
Linear shear stress distribution at the entrance & agrees with FE & - & \cmark & \xmark \\ \hline
\end{tabular}
\label{table:assumptions}
\end{table*}%
The table also shows the application of these assumptions in the slab method and FE. 
The three assumptions made by the FE are the assumption of plane-strain, which simplifies the problem into a 2D framework, the assumption of rigid rolls, and the use of an isotropic material model.  All the other assumptions were not imposed on the FE results, and therefore good comparison with the FE results can be seen as a justification of these other assumptions.  Plain-strain was assumed as otherwise the numerical cost of 3D FE simulations would have been intractable.  The assumption of plane-strain is common and experimentally verified away from the edges of wide sheets, although it would break down near the sheet edges.  To accurately predict edge phenomena such as bulging or wrinkling in these regions, therefore, the problem must be analyzed under a general 3D stress state, which was beyond the scope of this paper.  Similarly, non-rigid rolls (for example elastically deforming rolls) would add computational expense to the FE model; the assumption of rigid rolls is commonly regarded as valid for all but very thin foil rolling, where the flattening of the rolls limits the thinness of the sheet and requires a separate analysis.  In the mathematical model here, the roll shape ($h(z)$) is arbitrary and could be adapted to include roll flattening using analytical equations, such as Hitchcock's formula, when an appropriate contact condition in the flat segment is applied to model foil rolling.    
Isotropy is also a common assumption in models of rolling.  While the yield function could be replaced with an anisotropic function (e.g., Hill's~1948 yield criterion) with known anisotropy coefficients, the solution structure would largely remain the same, except with weighted stresses. However, this scenario is of limited scientific significance because it is the rolling process itself which induces anisotropy. To accurately capture this, the anisotropy would need to develop through the process, requiring a model for how deformation leads to anisotropy within the material, and introducing an additional layer of complexity to the modelling process.  The authors are not aware of studies modelling the development of anisotropy through the rolling process, even using FE analysis, and the prediction of the development of anisotropy by deformation remains an active area of research interest.  It is interesting to note in passing that exactly the same mathematical results are derived if the Tresca yield criterion is imposed instead of the von Mises yield criterion, provided the yield stress for the Tresca criterion is taken to be $\hat{\sigma}_Y = 2\hat\kappa$.  This is a consequence of the assumption of plane strain and the associated form taken by the stress tensor, so that we are always away from the kinks in the Tresca yield surface, and so suitably rescaling the yield stress (to account for Tresca being more conservative) results in the von Mises and Tresca criteria becoming identical in this case.

The other assumptions listed in table~\ref{table:assumptions} were not imposed on the FE results, and therefore good comparison with the FE results can be seen as a justification of these assumptions.  For example, the FE simulations were run forwards in time, and after the initial transients of the simulation had died away a steady state was reached from which the results were captured; if the modelling assumption of a steady state had not been valid, the FE simulations would not have been expected to reach a steady state.  Similarly, while elasticity plays a significant role both before and beyond the roll gap, particularly at the boundaries near the entrance and exit, the good agreement between the elasto-plastic FE and the rigid-plastic mathematical model demonstrates that the large pressures within the roll gap mean the behaviour is plastically-dominated within the roll gap, and the justification of neglecting elasticity there is valid.  Integrating elasto-plastic behaviour remains a challenging task, although without including elasticity, the models are unable to predict residual stress.  The plastic behaviour here is modelled as rigid (non-hardening) and isotropic.  Neglecting hardening has been shown here to have a negligible impact on the oscillations by comparing against hardening FE results, although including hardening within the model could be attempted in future research.

The systematic assumptions of a large aspect ratio, $\hat\ell/\hat h = 1/\delta$, and a small friction coefficient, $\mu$, which are common in practice, facilitate the asymptotic solution. For small aspect ratios, the solution remains valid, as demonstrated by the shear stress comparison in Figure~\ref{fig:shearcom} for a sheet with a length equal to its thickness. However, this requires a small reduction to satisfy our assumption of $h'$ being $O(\delta)$. The challenge in the small aspect ratio range, though, is that the entrance and exit regions constitute a significant portion of the overall length, making it less accurate to neglect elasticity. Employing large friction might result in the shear stress being of the same order as the normal stresses. Therefore, a different solution could be expected compared to the current prediction. 
The Coulomb law is used as the surface boundary condition, where the surface is assumed to slip along the roll throughout the roll gap unless a single point, which is the neutral point. However, we recognise that this assumption cannot always be justified and our FE results show a region of sticking around our predicted neutral point. 

The final set of assumptions relates to the condition of the sheet at the entrance.  Based on our assumption of rigid material outside the roll gap, if the sheet is to enter rigidly, the velocity profile must be uniform. However, unlike velocity, the assumption of rigid material outside the roll gap is insufficient to determine the force distribution at the entrance. Therefore, we proposed an assumed force distribution, which appears to be reasonable based on comparisons with FE simulations.
While homogeneous velocity and linear shear assumptions do not apply to all practical working conditions, they agree with the FE simulations of symmetric single-pass rolling, which do not rely on these assumptions. 
If the rolling process were asymmetric, or if previous rolling passes were to be modelled, different assumptions about the entrance condition of the material may be needed, which would be an interesting direction for future research.

\section{Conclusion}
\label{sec:conclusion}

A new semi-analytic model of the metal rolling process is presented, which, for the first time, successfully predicts the through-thickness stress and strain oscillations occurring in long, thin roll gaps.
Such oscillations are difficult to accurately compute using finite element (FE) analyses owing to the resolution needed to accurately resolve the oscillations.  The model is derived mathematically from the governing equations by systematically making simplifying assumptions; the assumptions made are summarized in table~\ref{table:assumptions}, and discussed in detail in section~\ref{sec:assumptions}.  Despite the simplifications, the model shows good agreement with high-resolution FE simulations while providing a significant computational speed-up.  The good agreement gives confidence not only in the model but also in the accuracy of the FE simulation compared against.  Furthermore, the asymptotic expansions and method of multiple scales used here allow the oscillations to be studied directly and automatically predict how they will change with changes in geometry.

Two length scales were introduced which enable the use of a multiple-scales asymptotic analysis: the large scale of the length of the roll gap, $\hat l$; and the small length scale of the initial sheet half-thickness, $\hat h_0$. 
The leading-order solution depends on the large length scale only and recreates the conventional pressure hill from slab analysis.  
The next-order correction is an order $O(\delta)$ smaller than the leading order, depends on both the large and small length scales, and provides a new prediction of through-thickness variation that shows an oscillatory pattern.  
This work is distinguished by this correction term, and generalises a previous asymptotic analysis by \citet{minton2016asymptotic} that incorrectly neglected these $O(\delta)$ correction terms.  
It should be noted that the shear stress is zero at leading order, and so this $O(\delta)$ ``correction term'' actually gives the leading-order behaviour of the shear stress.  
The first-order correction terms reveal two sets of waves for stress and velocity distribution varying on the short length scale of the sheet thickness.  
These are governed by a Burger's equation and an advection equation respectively, and can easily be solved numerically with suitable initial conditions at the roll-gap entrance. 
These corrections to the stress distribution have a minimal influence on the overall roll force and torque predictions, indicating that the accuracy of roll forces and torques are poor indications of the quality and accuracy of the entire simulation. 
Moreover, these fine details in the stress and velocity distribution will likely become significant when studying material properties, such as the hardening, development of anisotropy, or development of residual stress in the rolled sheet.

A set of rigorous FE simulations were performed using \Abaqus/Standard, as detailed by \citet{flanagan2024}.  
Significant work was needed for the FE simulation to be sufficiently accurate to give good agreement with the mathematical model.  
This highlights the usefulness of mathematical models for validating simulations.
In particular, commonly used mesh densities for rolling were found to be inadequate to capture the oscillatory pattern in the roll gap, and significantly higher mesh densities with 60 mesh points through-thickness were needed.  Moreover, convergence to a steady state and the avoidance of numerical oscillations typical of time-stepping FE simulations were given special attention.  The common initiation of rolling by pushing the workpiece into the roll gap was found to cause too many initial transients, and a steady state was only achieved numerically by using a two-step process with the rolls first pinching the workpiece and then beginning to roll.  These considerations minimized numerically-induced oscillations and allowed the use of standard available time-marching FE software (in this case \Abaqus), without the need to write a custom bespoke FE solver; the development of a custom-written bespoke FE solver designed to accurately and efficiently compute a steady state, for example by using a mixed Eulerian--Lagrangian formulation, might form interesting future research.
The asymptotic solution and FE simulations shown here demonstrate a high degree of agreement, validating both the asymptotic and FE results.
The solutions deviate in three places due to the simplifying assumptions of the mathematical model: the entrance and exit regions are sub-yield and are therefore governed by elasticity, which is neglected in the mathematical model; and the neutral point demonstrates sticking behaviour in the numerical simulations while only slipping friction was assumed in the mathematical model.  
These differences do not appear to affect either the quantitative predictions of the mathematical model elsewhere in the roll gap, nor the overall qualitative predictions of the model in general.

The mathematical model presented here offers a significantly better computational efficiency compared to FE simulations. 
The FE simulations for $\delta = 0.125$ used in this paper took 15.89 hours CPU time to run on a standard desktop computer, making them unsuitable for optimisation and real-time control. 
In contrast, the unoptimised \Matlab\ code evaluating the new mathematical model for the same $\delta$ took 3.45 seconds CPU time to run on a standard laptop which is more than 16,000 times faster compared to FE simulation. 

Further work could look at the entry and exit boundary conditions, likely by the inclusion of an elastic entrance and exit region. 
Similarly, including a sticking friction model near the neutral point would smooth the shock seen in the mathematical model but not the FE simulations. Appendix~\ref{appendix:neutral_point} would need to be revised in order to account for a sticking region of length $O(\delta)$.
Calculating higher-order terms, for example of $O(\delta^2)$, may or may not lead to more accurate results, and is unlikely to uncover new behaviour.
More mathematical techniques investigating slow, invariant, or centre manifolds may give either more insight or further accuracy~\citep{roberts1993invariant,valeryroy2002lubrication,roberts2015model}.
As mentioned above, incorporating roll flattening would allow the model to be applied to more extreme rolling cases such as foil rolling~\citep[extending][]{fleck1992cold}, and would involve adding an evolution equation to the roll shape $h(x)$ which is here taken to be arbitrary but known.  
The assumption of symmetry could also be relaxed to enable the prediction of asymmetric rolling~\citep[extending][]{minton2016asymptotic}, and temperature and recrystallisation effects could be included in order to model hot rolling processes.  
Including heating and heat flux in the model might allow the prediction of similar oscillatory patterns observed in the heat flux during rolling~\citep{olaogun2019heat}.  
Another extension might be to predict the stress field in other rolling processes, such as wire rolling~\citep{kazeminezhad2006calculation,vallellano2008analysis}.  
Finally, the present model's assumptions of perfect plasticity could be generalised to more realistic material models including strain-hardening and strain-rate-hardening, which is currently work in progress.  
Of course, all of these modifications would further complicate the model, making it harder to understand and interpret the fundamental mechanics; we therefore believe the present model represents a good balance between complexity and understandability, retaining most of the underlying simplicity of slab methods and introducing the wave-like behaviour needed for reasonable comparison with FE simulations. 

\section*{Acknowledgements}
ME gratefully acknowledges the support of a University of Warwick Chancellor's Scholarship.
EJB is grateful for the UKRI Future Leaders’ Fellowship funding (MR/V02261X/1) supporting this work.
FF is supported by Science Foundation Ireland Grant $\#$18/CRT/6049.
DOK is grateful for funding from the Science Foundation Ireland (21/FFP-P/10160).
ANOC is supported by the European Union, Science Foundation of Ireland and Lero, the Science Foundation Ireland Research Centre for Software, grants $\#$101028291
$\#$13/RC/2094 and $\#$SOWP2-TP0023, respectively.
For the purpose of open access, the authors have applied a Creative Commons Attribution (CC BY) licence to any Author Accepted Manuscript version arising from this submission.

A preliminary version of some parts of this work was presented at the 14th International Conference on the Technology of Plasticity, Mandelieu, France, September~2023~\citep{erfanian2023}.

\appendix
  \gdef\thesection{\Alph{section}}
  \makeatletter
  \renewcommand\@seccntformat[1]{\appendixname~\csname the#1\endcsname.\hspace{0.5em}}
  \makeatother

\section{The inner solution around the neutral point}\label{appendix:neutral_point}

As described in Section~\ref{sec:neutral_point}, in order to avoid a through-thickness discontinuity in $\sigma_{xy}^{(1)}$ as a result of $y\,\intd p^{(0)}\!/\intd z$ term (see change in the sign of shear stress through a vertical line at the neutral point in Figure \ref{fig:slab}), we carefully investigate here an inner region close to the neutral point. 
We introduce a new coordinate $X$ measuring distance near the neutral point,
\begin{equation} z=z_{\text{N}}+\delta X, \label{app:X}\end{equation}
where $z_{\text{N}}$ is the location of the neutral point obtained from the leading-order solution.  
We consider the neutral point to actually occur at $X = X_{\text{N}}$, where $X_{\text{N}}$ is an $O(\delta)$ correction to the location of the neutral point.
From~\eqref{app:X}, it can be seen that $X$ varies on the same length scale as the workpiece thickness, and so is comparable to the original $x$ coordinate in~\eqref{x}.  
Within the inner region, there is only one length scale, and so we do not need the multiple-scales variables $z$ and $n$; stresses are only functions of $X$ and $y$. 
Therefore, the governing equations are
\begin{subequations}\begin{gather}
\label{in-momentum} \pdr{\sigma_{{xx}_{\mathrm{inner}}}}{X}+  \pdr{\sigma_{{xy}_{\mathrm{inner}}}}{y} = 0
\qquad
\pdr{\sigma_{{yy}_{\mathrm{inner}}}}{y}+\pdr{\sigma_{{xy}_{\mathrm{inner}}}}{X} = 0 \\
\label{in-yield} \frac{1}{4}\left(\sigma_{{xx}_{\mathrm{inner}}} -\sigma_{{yy}_{\mathrm{inner}}}\right)^2+ \big( \sigma_{{xy}_{\mathrm{inner}}}\big)^2=1
\displaybreak[0]\\
\label{in-coulomb}\begin{aligned}
   \delta &\frac{dh}{dz} (\sigma_{{yy}_{\mathrm{inner}}} - \sigma_{{xx}_{\mathrm{inner}}})
   + \left(1-\delta ^{2}\!\left( \frac{dh}{dz}\right)^{\!2}\right)\!\sigma_{{xy}_{\mathrm{inner}}}
   \\& = \mp \delta\beta\left[
   \sigma_{{yy}_{\mathrm{inner}}}
   - 2\delta  \frac{dh}{dz}\sigma_{{xy}_{\mathrm{inner}}}
   + \delta ^{2}\left( \frac{dh}{dz}\right)^{\!2}\!\!\sigma_{{xx}_{\mathrm{inner}}}
   \right]
   \\&
   \qquad\text{at}\quad y=h(z).
  \end{aligned}\end{gather}\end{subequations}
The label ``inner'' distinguishes variables in the inner region from those in the outer region. 
At leading order, $\sigma_{{xy}_{\mathrm{inner}}}^{(0)}$ is zero similar to the outer region. 
From~\eqref{in-momentum} and~\eqref{in-yield} we find $\sigma_{{xx}_{\mathrm{inner}}}^{(0)}$, $\sigma_{{yy}_{\mathrm{inner}}}^{(0)}$ and $p_{\mathrm{inner}}^{(0)}$ are independent of $X$ and $y$, and are therefore constants,
 \begin{align} \label{normal0_inner}
 \sigma_{{xx}_{\mathrm{inner}}}^{(0)}=& 1-p^{(0)}_\mathrm{inner} & &\text{and}& 
\sigma_{{yy}_{\mathrm{inner}}}^{(0)}=& -1-p^{(0)}_\mathrm{inner}.   \end{align}

At $O(\delta)$, the yield condition \eqref{in-yield} becomes
\begin{align}
\sigma_{{xx}_{\mathrm{inner}}}^{(1)} -\sigma_{{yy}_{\mathrm{inner}}}^{(1)}&= 0& &\Rightarrow& \sigma_{{xx}_{\mathrm{inner}}}^{(1)} = \sigma_{{yy}_{\mathrm{inner}}}^{(1)}&= -p_{\mathrm{inner}}^{(1)} .
\end{align}
Substituting this into the local balance equations \eqref{in-momentum}, we obtain
\begin{align}
    -\pdr{p_{\mathrm{inner}}^{(1)}}{X}+ \pdr{\sigma_{{xy}_{\mathrm{inner}}}^{(1)}}{y} &= 0, &
-\pdr{p_{\mathrm{inner}}^{(1)}}{y}+ \pdr{\sigma_{{xy}_{\mathrm{inner}}}^{(1)}}{X} &=0.
\end{align}
The solution that satisfies the above differential equations is the wave-like solution
\begin{subequations}\label{in-p-1}\begin{align}
    p_{\mathrm{inner}}^{(1)}&= G\!\left(\frac{X}{h_{\text{N}}}+\frac{y}{h_{\text{N}}}\right)\!+G\!\left(\frac{X}{h_{\text{N}}}-\frac{y}{h_{\text{N}}}\right)\!,
    \\
    \sigma_{{xy}_{\mathrm{inner}}}^{(1)}&= G\!\left(\frac{X}{h_{\text{N}}}+\frac{y}{h_{\text{N}}}\right)\! - G\!\left(\frac{X}{h_{\text{N}}}-\frac{y}{h_{\text{N}}}\right)\!,
\end{align}\end{subequations}%
where $h_{\text{N}} = h(z_{\text{N}})$ is the half-thickness at the unperturbed neutral point and $G$ is an as-yet-unknown wave.
Substituting  $\sigma_{{xy}_{\mathrm{inner}}}^{(1)}$ into the friction equation \eqref{in-coulomb} at $O(\delta)$ gives the periodicity constraint on $G$ (with $\mp = \sgn(X-X_{\text{N}})$)
\begin{multline} \label{in-surface-1}
    - 2\left.\frac{\intd h}{\intd z}\right|_{z=z_N} \!\!\!\!\!\!+ \left[ G\left(\frac{X}{h_{\text{N}}}+1\right) - G\left(\frac{X}{h_{\text{N}}}-1\right) \right]
    \\
    \mp \beta (1+ p^{(0)}_\mathrm{inner}) =0.
\end{multline}  

This inner solution derived above should match with the previously derived solutions both to the left and to the right of the neutral point, as $|X|\to\infty$.  
Let us write $p^{(-)}$ and $p^{(+)}$ for the pressure calculated to the left ($-$) and to the right ($+$) of the neutral point, and similarly for $\sigma^{(\mp)}_{xy}$. 
Then the inner solution should match with these outer solutions when expanded in terms of $X$, so that, as $X \to \pm\infty$, we should have
\begin{subequations}
\begin{align}
p &= p^{(0\mp)}(y,z_{\text{N}}+\delta X)
\\\notag&\quad
+ \delta p^{(1\mp)}\big(n_{\text{N}}+X/h_{\text{N}}+O(\delta),y,z_{\text{N}}+\delta X\big) +O\big(\delta^2\big)
\\\notag
&= p^{(0)}(z_{\text{N}}) + \delta\!\left[A^{(\mp)}\!\left(\!n_{\text{N}}+\frac{X}{h_{\text{N}}}+\frac{y}{h_{\text{N}}},z_{\text{N}}\!\right)\!
\right.\\\notag&\left.\qquad\qquad\qquad\quad
+A^{(\mp)}\!\left(\!n_{\text{N}}+\frac{X}{h_{\text{N}}}-\frac{y}{h_{\text{N}}},z_{\text{N}}\!\right)\!
\right.\\\notag&\left.\qquad\qquad\qquad\quad
+ X\frac{dp^{(0\mp)}(z_{\text{N}})}{dz}\right] + O\big(\delta^2\big),
\\
\sigma_{xy} &= \sigma^{(1\mp)}_{xy}\big(n_{\text{N}} + X/h_{\text{N}} + O(\delta),y,z_{\text{N}}+\delta X\big) +  O(\delta^2)
\notag\\\notag
&= \delta\!\left[A^{(\mp)}\!\left(\!n_{\text{N}}+\frac{X}{h_{\text{N}}}+\frac{y}{h_{\text{N}}},z_{\text{N}}\!\right)\!
+y\frac{dp^{(0\mp)}(z_{\text{N}})}{dz}
\right.\\&\left.\qquad
-A^{(\mp)}\!\left(\!n_{\text{N}}+\frac{X}{h_{\text{N}}}-\frac{y}{h_{\text{N}}},z_{\text{N}}\!\right)
\right] + O\big(\delta^2\big),
\end{align} \label{out-expansion}%
\end{subequations}
where $p^{(1\mp)}$ and $\sigma_{xy}^{(1\mp)}$ are taken from equation~\eqref{eq:p_shear_first}, $p^{(0\mp)}$ satisfies~\eqref{eq:p_lead}, and $A^{(\mp)}(\xi,z)$ is the function $A(\xi,z)$ from the outer solution to the left ($-$) or right ($+$) of the neutral point.
The main reason for this appendix is to relate $A^{(+)}(\xi,z)$ and $A^{(-)}(\xi,z)$.  
Note that $p^{(0+)}(z_{\text{N}}) = p^{(0-)}(z_{\text{N}})$, as this is how the neutral point $z_{\text{N}}$ is chosen, but that $\intd p^{(0+)}\!/\intd z \neq \intd p^{(0-)}\!/\intd z$ at $z=z_{\text{N}}$ (see~\ref{eq:p_lead}).

At leading order, equation~\eqref{out-expansion} gives $p^{(0)}_\mathrm{inner} = p^{(0)}(z_{\text{N}})$, and confirms that $\sigma_{xy\mathrm{inner}}^{(0)} = 0$.  

We may exactly match $\sigma_{{xy}_{\mathrm{inner}}}^{(1)}$ and $p_{\mathrm{inner}}^{(1)}$ given in equation~\eqref{in-p-1} with the $O(\delta)$ part of equation~\eqref{out-expansion} with $(\mp)=(-)$ by taking
\begin{align}\label{app:G-}
&G(\xi)=A^{(-)}(n_{\text{N}}+\xi,z_{\text{N}}) + \frac{h_{\text{N}}}{2} \xi \frac{\intd p^{(0-)}}{\intd z}\Big| _{z=z_{\text{N}}}
\\\notag&\qquad\qquad
\text{for}\qquad \xi < \frac{X_N}{h_N}+1.
\end{align}
Since $A^{(-)}(n+1,z) = A^{(-)}(n-1,z)$ owing to~\eqref{eq:A:global}, and since $p^{(0-)}$ satisfies~\eqref{eq:p_shear_first}, this solution for $G(\xi)$ exactly satisfies the periodicity constraint~\eqref{in-surface-1} with $\mp=-$.
However, $\mp=-$ requires $X<X_N$, and hence $\xi < X_N/h_N+1$, as indicated in~\eqref{app:G-}.

Similarly, we may exactly match $\sigma_{{xy}_{\mathrm{inner}}}^{(1)}$ and $p_{\mathrm{inner}}^{(1)}$ given in equation~\eqref{in-p-1} with the $O(\delta)$ part of equation \eqref{out-expansion} with $(\mp)=(+)$ by taking
\begin{align}\label{app:G+}
&G(\xi)=A^{(+)}(n_{\text{N}}+\xi,z_{\text{N}}) + \frac{h_{\text{N}}}{2} \xi \frac{\intd p^{(0+)}}{\intd z}\Big| _{z=z_{\text{N}}}
\\\notag&\qquad\qquad
\text{for}\qquad \xi > \frac{X_N}{h_N}-1.
\end{align}
As before, since $A^{(+)}(n+1,z) = A^{(+)}(n-1,z)$ owing to~\eqref{eq:A:global}, this solution for $G(\xi)$ exactly satisfies the periodicity constraint~\eqref{in-surface-1} with $\mp=+$.
However, $\mp=+$ requires $X>X_N$, and hence $\xi > X_N/h_N-1$, as indicated in~\eqref{app:G+}.

In order to satisfy both~\eqref{app:G-} and~\eqref{app:G+}, for $X_N/h_N-1 < \xi < X_N/h_N+1$ we must have
\begin{subequations}\begin{align}
G(\xi)&=A^{(-)}(n_{\text{N}}+\xi,z_{\text{N}}) + \frac{h_{\text{N}}}{2} \xi \frac{\intd p^{(0-)}}{\intd z}\Big| _{z=z_{\text{N}}}
\\\notag
&= A^{(+)}(n_{\text{N}}+\xi,z_{\text{N}}) + \frac{h_{\text{N}}}{2} \xi \frac{\intd p^{(0+)}}{\intd z}\Big| _{z=z_{\text{N}}}
\\
\Rightarrow\quad
&A^{(+)}(n_{\text{N}}+\xi,z_{\text{N}})
\\\notag
&= A^{(-)}(n_{\text{N}}+\xi,z_{\text{N}}) + \frac{h_N}{2}\xi \left[ \frac{\intd p^{(0-)}}{\intd z}\Big| _{z=z_{\text{N}}}\!\!\!\!\! -  \frac{\intd p^{(0+)}}{\intd z}\Big| _{z=z_{\text{N}}}\right]
\\\notag
&= A^{(-)}(n_{\text{N}}+\xi,z_{\text{N}}) + \beta\xi (1+p^{(0)}(z_N)).
\label{A+atN}\end{align}\end{subequations}

In the outer region, the condition~\eqref{eq:A:global} results in $\int A^{(\mp)}(\xi)\, \mathrm{d} \xi = 0$. 
Therefore,
\begin{align}
   0&=\int A^{(+)}(\xi) \, \mathrm{d} \xi
\\\notag&
=\int\limits_{\mathrlap{X_{\text{N}}/h_{\text{N}}-1}}^{\mathrlap{X_{\text{N}}/h_{\text{N}}+1}} A^{(-)}(n_{\text{N}}+\xi)\,\mathrm{d} \xi
   + \int\limits_{\mathrlap{X_{\text{N}}/h_{\text{N}}-1}}^{\mathrlap{X_{\text{N}}/h_{\text{N}}+1}}\beta \xi\! \left(p^{(0)}(z_{\text{N}})+1\right) \mathrm{d} \xi
\\\notag&
   = 2\beta\!\left(p^{(0)}(z_{\text{N}})+1\right)\! \frac{X_{\text{N}}}{h_{\text{N}}}.
\end{align}
In order for this equality to be satisfied, we must take $X_{\text{N}}=0$. 
This means there is no $O(\delta)$ correction to the location of the neutral point, and the neutral point is located at $z=z_{\text{N}}$, found from equation \eqref{eq:p_lead} for $p^{(0)}$ in the outer region.  
Equation~\eqref{A+atN} thus gives the connection between the solutions to the left and right of the neutral point given in~\eqref{eq:Adiscontinuity}.

\section{Solving for the second-order stresses} \label{App:sigma_xy_second}

In this appendix, the solution is continued to this order of correction with the goal of finding the unknown parameters $A(\xi,z)$ and $D(z)$ from the previous order. 
From the yield function \eqref{eq:yield}, at $O(\delta^2)$, and substituting the known variables from the previous orders, we find that
\begin{align} \label{eq:sigma_second} &\sigma_{xx}^{(2)} = -p^{(2)} - \frac{1}{2} \sigma_{xy}^{(1)} \strut^2  &\text{and}&  &\sigma_{yy}^{(2)} = -p^{(2)} + \frac{1}{2} \sigma_{xy}^{(1)} \strut^2.
\end{align}
 At $O(\delta^2)$, the local balance equations~\eqref{eq:mom1} and~\eqref{eq:mom2} are
\begin{subequations}\label{eq:mom_second}\begin{align}  \pdr{\sigma_{xx}^{(1)}}{z}+  \frac{1}{h}\pdr{\sigma_{xx}^{(2)}}{n} + \pdr{\sigma_{xy}^{(2)}}{y}&= 0, \\
 \pdr{\sigma_{yy}^{(2)}}{y}+\pdr{\sigma_{xy}^{(1)}}{z} +\frac{1}{h}\pdr{\sigma_{xy}^{(2)}}{n}&=0. \end{align}\end{subequations}%
 By substituting in $\sigma_{xx}^{(2)}$ and $\sigma_{yy}^{(2)}$ from equation \eqref{eq:sigma_second}, and $\sigma_{xx}^{(1)}$ from~\eqref{eq:sigma_first}, we have,
\begin{subequations}\label{eq:mom_second_sim}\begin{align}
  \pdr{\sigma_{xy}^{(2)}}{y} -  \frac{1}{h}\pdr{p^{(2)}}{n}&= \pdr{p^{(1)}}{z} + \frac{1}{2h}\pdr{\sigma_{xy}^{(1)} \strut^2}{n},
\\
 \pdr{p^{(2)}}{y} - \frac{1}{h}\pdr{\sigma_{xy}^{(2)}}{n}&= \pdr{\sigma_{xy}^{(1)}}{z} + \frac{1}{2}\pdr{\sigma_{xy}^{(1)} \strut^2}{y} . 
\end{align}\end{subequations}%
These two equations give wave equations for $p^{(2)}$ and $\sigma_{xy}^{(2)}$, forced by $p^{(1)}$ and $\sigma_{xy}^{(1)}$.
By substituting  $p^{(1)}$ and  $\sigma_{xy}^{(1)}$ from equation \eqref{eq:p_shear_first} into~\eqref{eq:mom_second_sim}, we arrive at
\begin{subequations}\begin{align}\label{mom1_ext_second}
&\pdr{\sigma_{xy}^{(2)}}{y} - \frac{1}{h(z)}\pdr{p^{(2)}}{n}
\\\notag&\quad
=\left[ \pdr{A}{z}\left(n+\frac{y}{h}\right) + \pdr{A}{z}\left(n-\frac{y}{h}\right) \right]  - \frac{dD(z)}{dz}
\\\notag&\qquad
+ \frac{1}{2h} \pdr{}{n} \left[ A\left(n+\frac{y}{h}\right)^2 + A\left(n-\frac{y}{h}\right)^2
\right.\\\notag&\qquad\qquad\qquad\left.
- 2A\left(n+\frac{y}{h}\right)A\left(n-\frac{y}{h}\right)\right]
\\\notag&\qquad
+ \frac{y}{h} \left[ A'\left(n+\frac{y}{h}\right) - A'\left(n-\frac{y}{h}\right) \right] \Big( \frac{dp^{(0)}}{dz} - \frac{\intd h/ \intd z}{h} \Big)
\displaybreak[0]
\\[1ex]
\label{mom2_ext_second}
&\pdr{p^{(2)}}{y} - \frac{1}{h}\pdr{\sigma_{xy}^{(2)}}{n}
\\\notag&\quad
= \left[ \pdr{A}{z}\left(n+\frac{y}{h}\right) - \pdr{A}{z}\left(n-\frac{y}{h}\right) \right]  \\\notag&\qquad
+ \frac{1}{2} \pdr{}{y} \left[ A\left(n+\frac{y}{h}\right)^2 + A\left(n-\frac{y}{h}\right)^2
\right.\\\notag&\qquad\qquad\qquad\left.
- 2A\left(n+\frac{y}{h}\right)A\left(n-\frac{y}{h}\right) \right]
\\\notag&\qquad
+ \frac{y}{h} \left[ A'\left(n+\frac{y}{h}\right) + A'\left(n-\frac{y}{h}\right) \right] \Big( \frac{dp^{(0)}}{dz} - \frac{\intd h/ \intd z}{h} \Big)
\\\notag&\qquad
+ y \left( \left(\frac{dp^{(0)}}{dz}\right)^2 + \ppdr{p^{(0)}}{z} \right)
\\\notag&\qquad
+ \frac{dp^{(0)}}{dz} \left[ A\left(n+\frac{y}{h}\right) - A\left(n-\frac{y}{h}\right) \right], 
\end{align}\end{subequations}%
where prime denoting $\partial/\partial \xi$, and $A(n \pm ({y}/{h}),z)$ is abbreviated to $A(n \pm ({y}/{h}))$ for brevity while the dependence on $z$ is still acknowledged. 
We solve this as a system of two coupled equations by finding particular integrals for each of the forcing terms of the right-hand-side.
Ultimately, we find the solution for $\sigma_{xy}^{(2)}$ and $p^{(2)}$ as
\begin{subequations}\begin{align} 
\sigma^{(2)}_{xy} = &\, M\!\left(n+\frac{y}{h}\right) - M\!\left(n-\frac{y}{h}\right)
\\\notag&
+ y \!\left[ \pdr{A}{z}\!\left(n+\frac{y}{h}\right) + \pdr{A}{z}\!\left(n-\frac{y}{h}\right) \right]
\\\notag&
+ \frac{y^2}{2h} \Big( \frac{dp^{(0)}}{dz} - \frac{\intd h/ \intd z}{h} \Big)\!\!\left[ A'\!\left(n+\frac{y}{h}\right)\! - A'\!\left(n-\frac{y}{h}\right) \right]
\\\notag&
+ \frac{y}{2}  \frac{dp^{(0)}}{dz} \!\left[ A\!\left(n+\frac{y}{h}\right) + A\!\left(n-\frac{y}{h}\right) \right]
\\\notag& 
+ \frac{1}{2} \!\left[  A'\!\left(n+\frac{y}{h}\right)\! \intA\!\left(n-\frac{y}{h}\right)- \intA\!\left(n+\frac{y}{h}\right)\! A'\!\left(n-\frac{y}{h}\right) \right]
\\\notag&
+ \frac{y}{2h} \pdr{}{n} \!\left[ A\!\left(n+\frac{y}{h}\right) \strut ^ 2 + A\!\left(n-\frac{y}{h}\right) \strut ^ 2 \right]  - y\frac{dD(z)}{dz},
\displaybreak[0]\\[1ex]
p^{(2)} = &\, M\!\left(n+\frac{y}{h}\right) + M\!\left(n-\frac{y}{h}\right)
\\\notag& 
+ y \!\left[ \pdr{A}{z}\!\left(n+\frac{y}{h}\right) - \pdr{A}{z}\!\left(n-\frac{y}{h}\right) \right]
\\\notag&
+ \frac{y^2}{2h} \Big( \frac{dp^{(0)}}{dz} - \frac{\intd h/ \intd z}{h} \Big) \!\left[ A'\!\left(n+\frac{y}{h}\right)\! + A'\!\left(n-\frac{y}{h}\right) \right]
\\\notag& 
+ \frac{y}{2}  \frac{dp^{(0)}}{dz} \!\left[ A\!\left(n+\frac{y}{h}\right) - A\!\left(n-\frac{y}{h}\right)\! \right]
\\\notag& 
+ \frac{h}{2} \frac{dp^{(0)}}{dz} \!\left[ \intA(n+\frac{y}{h}) + \intA(n-\frac{y}{h}) \right]
\\\notag&
+ \frac{1}{2} \!\left[  A'\!\left(n+\frac{y}{h}\right)\! \intA\!\left(n-\frac{y}{h}\right) + \intA\!\left(n+\frac{y}{h}\right)\! A'\!\left(n-\frac{y}{h}\right) \right]
\\\notag& + \frac{y}{2h} \pdr{}{n} \!\left[ A\!\left(n+\frac{y}{h}\right)^2 - A\!\left(n-\frac{y}{h}\right)^2 \right] 
\\\notag& 
+ \frac{y^2}{2}\!\left(\frac{d^2p^{(0)}(z)}{dz^2} + \!\left(\frac{dp^{(0)}(z)}{dz}\right)^{\!2} \right),
\end{align}\end{subequations}%
where the term $M\left(n+({y}/{h})\right) - M\left(n-({y}/{h})\right)$ represents the complementary solution, which is an as-yet-unknown wave, and $\intA$ is the integral of $A$ such that $\intA' = A$.

Substituting this solution into the friction equation at this order, $\sigma_{xy}^{(2)} \mp \beta p^{(1)} = 0$,  results in
\begin{align}
0 =& M(n+1,z) - M(n-1,z)
\\\notag&
+ h \left[ \pdr{A(n+1,z)}{z} + \pdr{A(n-1,z)}{z} \right]
\\\notag&
+ \frac{h}{2} \Big( \frac{dp^{(0)}}{dz} - \frac{1}{z}\frac{\intd h}{\intd z} \Big)\!  \big[ A'(n+1,z) - A'(n-1,z) \big]
\\\notag&
+ \frac{h}{2}  \frac{dp^{(0)}}{dz} \big[ A(n+1,z) + A(n-1,z) \big]
\\\notag&
 + \frac{1}{2}  \big[  A'(n+1,z) \intA(n-1,z)-  A'(n-1,z)\intA(n+1,z) \big]
\\\notag&
 + \frac{1}{2} \pdr{}{n} \left[ A(n+1,z) \strut ^ 2 + A(n-1,z) \strut ^ 2 \right]
 - h\pdr{D(z)}{z}
\\\notag&
 \mp \beta \big[A(n+1,z) + A(n-1,z) + D(z) \big].
\end {align}

This equation can be significantly simplified by using the condition~\eqref{eq:A:global}, to give
\begin{multline} \label{App:A}
  2h\pdr{A(n+1,z)}{z}   + \pdr{}{n} \Big( A(n+1,z) \strut ^ 2 \Big)
  \\\
  +\left(h  \frac{dp^{(0)}}{dz} \mp 2 \beta 
 \right)A(n+1,z)
   - h\frac{dD(z)}{dz} \mp \beta D(z)
\\
   = -[M(n+1,z) - M(n-1,z)].
\end{multline}
Where the term $M(n+1,z) - M(n-1,z)$ is the wave-equation solution for $\sigma_{xy}^{(2)}$ to the homogeneous version of the coupled problem in~\eqref{eq:mom_second_sim}.  
This is the equation quoted in~\eqref{eq:A}.

\subsection*{Rearranging the second order stresses to get Burger's equation}

As discussed following~\eqref{eq:A}, we do not need to calculate $M$, as we need only require that $M$ be bounded in $n$, and hence that the right-hand side of equation \eqref{App:A} be zero.
Integrating~\eqref{App:A} between $n=-2$ and $n=0$ and imposing~\eqref{eq:A:global} then gives
\begin{equation}
     -h\frac{\intd D}{\intd z} \mp \beta D(z) =0.
\end{equation} 
Since $D(0)=0$ from~\eqref{eq:D_ic}, we conclude that $D(z) \equiv 0$. 
With this, equation~\eqref{App:A} then gives
\begin{equation} \label{simplified}
  2h\pdr{A(\xi,z)}{z}   + \pdr{}{\xi} \Big(\! A(\xi,z) \strut ^ 2 \Big) +\!\left(\! h  \frac{dp^{(0)}}{dz}  \mp 2 \beta \!\right)\!A(\xi,z) = 0. \end{equation} 
The factor $h\,\intd p^{(0)}\!/\intd z$ can be replaced with its equivalent from~\eqref{eq:p_lead},
\begin{multline} \label{simplified_2}
  \pdr{A(\xi,z)}{z}   + \frac{1}{2h} \pdr{}{\xi} \Big(\! A(\xi,z) \strut ^ 2 \Big)
  \\
  - \left(\! \mp \frac{\beta}{2h}(p^{(0)}-1) - \frac{\intd h/ \intd z}{h}\right)\!A(\xi,z) = 0. \end{multline}
We now aim to combine the first and third terms of~\eqref{simplified_2} into a single $z$-derivative.  
To this end, we first rewrite the factor multiplying $A$ in the final term as a $z$-derivative by defining,
\begin{equation} \alpha_1(z) = \exp\! \Bigg\{\! \int ^ z _0 \Big(\! \mp \frac{\beta }{2h(\Tilde{z})}(p^{(0)} (\Tilde{z})-1) \Big) d\Tilde{z} \Bigg\}, \end{equation}
Differentiating $\log(\alpha_1/h)$ with respect to $z$ then shows that
\begin{equation}
\Big( \mp \frac{\beta}{2h}(p^{(0)}-1) - \frac{\intd h/ \intd z}{h}\Big) = \frac{\pdr{}{z} (\alpha_1 / h)}{(\alpha_1 / h)}. 
\end{equation}
Replacing this into the original equation \eqref{simplified_2}, this equation can be rewritten as
\begin{equation} \label{Burgers}
\pdr{A(\xi)}{z} -  \frac{\pdr{}{z} (\alpha_1 / h)}{(\alpha_1 / h)}  A(\xi)  + \frac{1}{2h} \pdr{}{\xi} \Big( A(\xi) \strut ^ 2 \Big)  = 0. \end{equation}
By dividing~\eqref{Burgers} by $\alpha_1 / h$, the first two terms give a $z$ derivative of $\frac{A(\xi)}{(\alpha_1 / h)}$. 
Thus,
\begin{equation}\pdr{}{z} \left(\frac{A(\xi)}{(\alpha_1 / h)}\right) + \frac{\alpha_1}{2 h^2} \pdr{}{\xi} \left(\left( \frac{A(\xi)}{(\alpha_1 / h)}\right)^2 \right) =0.\end{equation}
Finally, with the following change of variable, Burger's equation is obtained,
\begin{multline}\label{App:Burgers_total}
T  = \int^z \frac{\alpha_1}{h(\bar z)\strut^2}\ \mathrm{d} \bar z
\quad\text{and}\quad 
\omega\big(\xi,T(z)\big) = \frac{A(\xi,z)}{(\alpha_1 / h)}
\\
\Rightarrow\quad
\pdr{}{T} (\omega) + \frac{1}{2}  \pdr{}{\xi} \Big( \omega \strut ^ 2 \Big) = 0.
\end{multline}
This is the equation as quoted in~\eqref{eq:Burgers_total}.

\section{Solving for the second-order velocities} \label{App:v_second}

Here we solve equations~\eqref{eq:vel_second} which describe the velocity at $O(\delta^2)$.
This does not completely resolve the velocity, as it will be determined only up to an unknown function $N(\xi,z)$ which would be determined at the next order, $O(\delta^3)$.
However, as for the second order stresses, it does gives us a secularity condition for the previously unknown function $B(\xi,z)$ which occurs in $v^{(1)}$, and that allows us to completely determine $v^{(1)}$.

We begin with equation~\eqref{eq:vel_second}, given by
\begin{subequations}\label{app:vel_second}\begin{align}
\pdr{v^{(2)}}{y}+\frac{1}{h}\pdr{u^{(2)}}{n}&=-\pdr{u^{(1)}}{z},\\
\pdr{u^{(2)}}{y}+\frac{1}{h}\pdr{v^{(2)}}{n}&=-\pdr{v^{(1)}}{z} + 2\lambda^{(1)} \sigma^{(1)}_{xy}.
\end{align}\end{subequations}
Substituting the solutions obtained for $u^{(1)}$, $v^{(1)}$, $\lambda^{(1)}$, and $\sigma_{xy} ^{(1)}$ from equation \eqref{eq:vel_first}, \eqref{eq:lambda_first}, and \eqref{eq:p_shear_first} into~\eqref{app:vel_second} yields
\begin{subequations}\begin{align}
\label{flow_exp_second}
&\pdr{u^{(2)}}{y}+\frac{1}{h}\pdr{v^{(2)}}{n}
\\\notag&\quad
= \left[ \pdr{B}{z}\!\left(n+\frac{y}{h}\right) -  \pdr{B}{z}\!\left(n-\frac{y}{h}\right)\right]
\\\notag&\qquad
- 2\frac{\intd h/\intd z}{h^2} \left[ A \!\left(n+\frac{y}{h}\right) -  A \!\left(n-\frac{y}{h}\right)\right]
\\\notag&\qquad
+y \!\left(\frac{2}{h} \frac{\intd p^{(0)}}{\intd z}-\frac{\intd h/\intd z}{h^2}\right)\! \left[ B'\!\left(n+\frac{y}{h}\right) +  B'\!\left(n-\frac{y}{h}\right)\right]
\\\notag&\qquad
+ \frac{2}{h} \left[ A \!\left(n+\frac{y}{h}\right) -  A \!\left(n-\frac{y}{h}\right)\right]\! \left[ B'\!\left(n+\frac{y}{h}\right) +  B'\!\left(n-\frac{y}{h}\right)\right]
\\\notag&\qquad
- y \!\left( \frac{h (\intd^2 h/\intd z^2) -2(\intd h/\intd z)^2}{h^3} + 2\frac{ \intd h/\intd z}{h^2} \frac{\intd p^{(0)}}{\intd z} \right) 
\displaybreak[0]\\[1ex]\label{conti_exp_second}
&\pdr{v^{(2)}}{y}+\frac{1}{h}\pdr{u^{(2)}}{n}
\\\notag&\quad
= -\left[ \pdr{B}{z}\!\left(n+\frac{y}{h}\right) +  \pdr{B}{z}\!\left(n-\frac{y}{h}\right)\right]
\\\notag&\qquad
+y \frac{\intd h/\intd z}{h^2}\left[ B'\!\left(n+\frac{y}{h}\right) -  B'\!\left(n-\frac{y}{h}\right)\right],
\end{align} \label{app:velocity}\end{subequations}%
where $A(\xi,z)$ is abbreviated to $A(\xi)$ for brevity, and similarly for $B(\xi)$. 
These equations form a coupled wave equation for $u^{(2)}$ and $v^{(2)}$, forced by $A(\xi)$ and $B(\xi)$.   
Solving them as a system of coupled equations, by finding particular integrals for each of the forcing terms on the right-hand-side, ultimately, results in
\begin{subequations}\begin{align}
v^{(2)} = &
- \left[ N \!\left(n+\frac{y}{h}\right) - N \!\left(n-\frac{y}{h}\right) \right]
\\\notag&
- y \left[ \pdr{B}{z}\!\left(n+\frac{y}{h}\right) +  \pdr{B}{z}\!\left(n-\frac{y}{h}\right) \right]
\\\notag&   
  + \frac{\intd h/\intd z}{h^2} y \left[ A \!\left(n+\frac{y}{h}\right) +  A \!\left(n-\frac{y}{h}\right) \right]
\\\notag&   
    +  y  \frac{1}{2} \frac{\intd p^{(0)}}{\intd z} \left[ B\!\left(n+\frac{y}{h}\right) +  B\!\left(n-\frac{y}{h}\right) \right]
\\\notag&   
 - \frac{1}{2}y^2 \!\left(\frac{1}{h} \frac{\intd p^{(0)}}{\intd z}-\frac{\intd h/\intd z}{h^2}\right) \left[B'\!\left(n+\frac{y}{h}\right) -  B'\!\left(n-\frac{y}{h}\right) \right]
\\\notag&   
  - \frac{y}{h}  \left[ B'\!\left(n+\frac{y}{h}\right)A \!\left(n+\frac{y}{h}\right) +B'\!\left(n-\frac{y}{h}\right)A \!\left(n-\frac{y}{h}\right) \right]
\\\notag&   
  - \frac{1}{2}  \left[ B\!\left(n+\frac{y}{h}\right) A \!\left(n-\frac{y}{h}\right) + B'\!\left(n+\frac{y}{h}\right) \intA \!\left(n-\frac{y}{h}\right) \right]
\\\notag&   
  + \frac{1}{2}  \left[ B\!\left(n-\frac{y}{h}\right) A \!\left(n+\frac{y}{h}\right) + B'\!\left(n-\frac{y}{h}\right) \intA \!\left(n+\frac{y}{h}\right) \right],
\displaybreak[0]\\[1ex]
u^{(2)} = &
 \left[ N \!\left(n+\frac{y}{h}\right) + N \!\left(n-\frac{y}{h}\right) \right]
\\\notag&   
 + y \left[ \pdr{B}{z}\!\left(n+\frac{y}{h}\right) -  \pdr{B}{z}\!\left(n-\frac{y}{h}\right) \right]
 \\\notag&     
  - \frac{\intd h/\intd z}{h^2} y \left[ A \!\left(n+\frac{y}{h}\right) -  A \!\left(n-\frac{y}{h}\right) \right]
\\\notag&   
  +  y  \frac{1}{2} \frac{\intd p^{(0)}}{\intd z} \left[ B\!\left(n+\frac{y}{h}\right) -  B\!\left(n-\frac{y}{h}\right) \right]
\\\notag&   
 + \frac{1}{2}y^2 \!\left(\frac{1}{h} \frac{dp^{(0)}}{dz}-\frac{\intd h/\intd z}{h^2}\right) \left[ B'\!\left(n+\frac{y}{h}\right) +  B'\!\left(n-\frac{y}{h}\right) \right]
\\\notag&   
  + \frac{y}{h}  \left[ B'\!\left(n+\frac{y}{h}\right)A \!\left(n+\frac{y}{h}\right) - B'\!\left(n-\frac{y}{h}\right)A \!\left(n-\frac{y}{h}\right) \right]
\\\notag&   
  - \frac{1}{2}  \left[ B\!\left(n+\frac{y}{h}\right) A \!\left(n-\frac{y}{h}\right) - B'\!\left(n+\frac{y}{h}\right) \intA \!\left(n-\frac{y}{h}\right) \right]
\\\notag&   
  - \frac{1}{2}  \left[ B\!\left(n-\frac{y}{h}\right) A \!\left(n+\frac{y}{h}\right) - B'\!\left(n-\frac{y}{h}\right) \intA \!\left(n+\frac{y}{h}\right) \right]
\\\notag&   
  -\frac{1}{2}y^2 \!\left( \frac{h (\intd^2 h/\intd z^2) -2(\intd h/\intd z)^2}{h^3} + 2\frac{ \intd h/\intd z}{h^2} \frac{dp^{(0)}}{dz} \right)
\\\notag&   
  - \frac{\intd h/\intd z}{h} \left[ \intA \!\left(n+\frac{y}{h}\right) +  \intA \!\left(n-\frac{y}{h}\right) \right]
\\\notag&   
  - \frac{h}{2} \frac{\intd p^{(0)}}{\intd z} \left[ \intB \!\left(n+\frac{y}{h}\right) +  \intB \!\left(n-\frac{y}{h}\right) \right]
\\\notag&   
  + \!\left[ \!\left(\!B'\!\!\left(\!n+\frac{y}{h}\right)\!A \!\left(\!n+\frac{y}{h}\right)\! \right)^{\!\wedge}\! + \!\left (\!B'\!\!\left(\!n-\frac{y}{h}\right)\!A \!\left(\!n-\frac{y}{h}\right)\! \right)^ {\!\wedge} \right]\! .
\end{align}\end{subequations}%
where the term $-\left[ N\left(n+({y}/{h})\right) - N\left(n-({y}/{h})\right)\right]$ represents the complementary solution, which is an as-yet-unknown wave, and $\intB$ is the integral of $A$ such that $\intB' = B$.

Replacing $v^{(2)}$ into the boundary condition~\eqref{eq:boundary2 velocity} at the second order, $v^{(2)} = (\intd h/\intd z) u^{(1)}$, results in
\begin{align}
&\frac{dh}{dz} \Big( B(n+1) +  B(n-1)\Big)
\\\notag\quad&   
=
- \Big( N (n+1) - N (n-1)\Big)
\\\notag\qquad&   
+ \frac{1}{h}\frac{\intd h}{\intd z} \Big( A (n+1) +  A (n-1)\Big)
\\\notag\qquad&   
-h \Big( \pdr{B(n+1)}{z}  +  \pdr{B(n-1)}{z} \Big)
\\\notag\qquad&   
+ \frac{1}{2}h^2 \!\left(\frac{1}{h} \frac{\intd p^{(0)}}{\intd z}-\frac{\intd h/\intd z}{h^2}\right)\!\Big(\! -B' (n+1) +  B' (n-1)\Big)
\\\notag\qquad&   
+ \frac{h}{2} \!\left(\frac{\intd p^{(0)}}{\intd z}\right) \Big( B (n+1) +  B (n-1)\Big)
\\\notag\qquad&   
-  \frac{1}{2} \Big(\!B(n+ 1) A(n- 1) + B'(n+ 1) \intA (n- 1)\! \Big)
\\\notag\qquad&   
- \Big(\!B'(n+ 1)A(n+ 1) +B'(n- 1)A(n- 1)\! \Big)\!
\\\notag\qquad&   
+ \frac{1}{2} \Big(\!B(n- 1) A(n+ 1) + B'(n- 1) \intA (n+ 1) \!\Big)
\end{align}
This equation can be extensively simplified on the surface utilising the periodicity of functions $A$ and $B$, using equations~\eqref{eq:A:global} and~\eqref{eq:Bperiodic}, to obtain
\begin{multline}\label{App:simplified_3}
\pdr{B(n+1,z)}{z}
- \frac{\intd h/\intd z}{h^2}A(n+1,z)
\\
- \left(\frac{1}{2}\frac{\intd p^{(0)}}{\intd z} - \frac{\intd h/\intd z}{h}\right)\!B(n+1,z)
\\
+ \frac{1}{h}B'(n+1,z)A(n+1,z)
\\
= -\frac{1}{h} \!\left( N (n+1,z) -  N (n-1,z)\right).   
\end{multline}
This is expressions~\eqref{eq:simplified_3} quoted in Section~\ref{v2}.

\subsection*{Rearranging the second order velocities to get an advection equation} \label{App:advection}

As argued following~\eqref{eq:simplified_3}, avoiding $N$ growing as a function of $n$ gives the secularity condition $N(n+1,z) = N(n-1,z)$, meaning the right-hand-side of~\eqref{App:simplified_3} is zero.
This equation can then be further simplified by defining $\alpha_2(z) = \exp\big\{\frac{1}{2}p^{(0)}\big\}$, to give,
\begin{equation} \label{App:advection_A} \pdr{}{z} \left(\frac{B(\xi,z)}{(\alpha_2 / h)}\right) + \frac{A(\xi,z)}{h}  \pdr{}{\xi} \left(\frac{B(\xi,z)}{(\alpha_2 / h)}\right) = \frac{1}{h\mkern 1mu \alpha_2} \frac{\intd h}{\intd z}A(\xi,z). 
\end{equation}
This is a forced advection equation with a velocity of $A(\xi,z)/h$. 
It is, however, more convenient to solve the equation if the excitation term on the right hand side can be included inside the derivatives.
Moreover, if the advection speed is the same as Burger's equation for the stresses~\eqref{eq:Burger}, both equations share the same characteristics, again aiding numerical solution.
This is the aim of the rearrangement detailed here.

We start transforming equation \eqref{App:advection_A} by multiplying it by $h^2 \!/ \alpha_1$, giving
\begin{equation} \label{eq:advection_B}
\frac{h^2}{\alpha_1} \pdr{}{z} \Big(\frac{B}{(\alpha_2 / h)}\Big) + \omega  \pdr{}{\xi} \Big(\frac{B}{(\alpha_2 / h)}\Big) - \frac{\intd h/ \intd z}{\alpha_2}\omega = 0, \end{equation} 
where $\omega$ is as was defined for the stresses and is given by~\eqref{App:Burgers_total}.
To remove the excitation term, we look at the multiple of $\omega$ which solves equation \eqref{eq:advection_B}. 
Calling this factor $Q(z)$, in this case we can write
\begin{multline} \label{Qw} \frac{h^2}{\alpha_1}\pdr{}{z} \Big(\!Q \omega\!\Big) + \omega \pdr{}{\xi} \Big(\!Q\omega \!\Big)- \frac{\omega}{\alpha_2}\frac{\intd h}{\intd z}
\\
= 
 \frac{h^2}{\alpha_1} \omega \frac{dQ}{dz} + Q\! \left[ \frac{h^2}{\alpha_1} \pdr{\omega}{z}  + \frac{1}{2}  \pdr{}{\xi} \Big( \omega^2 \Big) \right]-\frac{\omega}{\alpha_2}\frac{\intd h}{\intd z}
 \;=\; 0.\end{multline}
The term in the bracket is zero siunce $\omega$ satisfies Burger's equation \eqref{App:Burgers_total}. 
The remaining terms find $Q(z)$ as
\begin{align} \label{eq:QW}
    \frac{h^2}{\alpha_1(z)}\frac{dQ}{dz} &= \frac{\intd h/ \intd z}{\alpha_2(z)} &&\Rightarrow& Q(z) &= \int ^z \frac{\alpha_1(\bar z)}{\alpha_2 (\bar z)}\frac{\intd h(\bar z)/\intd(\bar z)}{h(\bar z)^2} d \bar z. 
\end{align}
Subtracting \eqref{eq:QW} from \eqref{eq:advection_B} results in
\begin{multline}
\frac{h^2}{\alpha_1}\pdr{}{z} \Big(\frac{B(\xi,z)}{(\alpha_2 / h)}-Q(z)\omega(\xi,z)\Big)
\\
+ \omega(\xi,z)  \pdr{}{\xi} \Big(\frac{B(\xi,z)}{(\alpha_2 / h)}-Q(z)\omega(\xi,z)\Big) = 0. \end{multline}
With the same change of domain $z \mapsto T$ as \eqref{App:Burgers_total}, we finally have,
\begin{equation} \label{advection}
     \pdr{}{T} \Big(\frac{B}{(\alpha_2 / h)}-Q\omega\Big) + \omega  \pdr{}{\xi} \Big(\frac{B}{(\alpha_2 / h)}-Q\omega\Big)=0,
\end{equation}
which is the expression given in equation~\eqref{eq:advection_w}.

\bibliography{main} 

\end{document}